\documentclass[11pt,preprint]{aastex} 

\usepackage{natbib}
\received{}
\accepted{}
\shorttitle{Parallax to SY Aur}
\shortauthors{Riess et al.}

\newcommand{\bq}{\begin{equation}} 
\newcommand{\eq}{\end{equation}}   

\newcommand{\beq}{\begin{equation}}
\newcommand{\eeq}{\end{equation}}
\newcommand{\beqa}{\begin{eqnarray}}
\newcommand{\eeqa}{\end{eqnarray}}

\newcommand{\PL}{$P\hbox{--}L$}

\newcommand{\muas}{\hbox{$\, \mu\rm as$}}
\newcommand{\mpix}{\hbox{$\, \rm mpix$}}
\newcommand{\mas}{\hbox{$\, \rm mas$}}
\newcommand{\kpc}{\hbox{$\, \rm kpc$}}
\newcommand{\Filter}[1]{\hbox{\sl #1}}

\begin{document} 

\title{Parallax Beyond a Kiloparsec from Spatially Scanning \\ 
the Wide Field Camera 3 on the Hubble Space Telescope\altaffilmark{1}}

\author{Adam G. Riess\altaffilmark{2,3}, Stefano
Casertano\altaffilmark{3,2}, Jay Anderson\altaffilmark{3}, John
MacKenty\altaffilmark{3}, and Alexei V. Filippenko\altaffilmark{4}}

\altaffiltext{1}{Based on observations with the NASA/ESA {\it Hubble Space
  Telescope}, obtained at the Space Telescope Science Institute, which is
  operated by AURA, Inc., under NASA contract NAS 5-26555.}
\altaffiltext{2}{Department of Physics and Astronomy, Johns Hopkins
  University, Baltimore, MD 21218.}
\altaffiltext{3}{Space Telescope Science Institute, 3700 San Martin
  Drive, Baltimore, MD 21218; ariess@stsci.edu.}
\altaffiltext{4}{Department of Astronomy, University of California,
  Berkeley, CA 94720-3411.}
  
\begin{abstract}

  We use a newly developed observing mode on the {\it Hubble Space
  Telescope (HST)} and Wide Field Camera 3 (WFC3), spatial scanning,
  to increase source sampling a thousand-fold and measure changes in
  source positions to a precision of $ 20\hbox{--}40 \muas $, more than an order of
  magnitude better than attainable in pointed observations.  This
  observing mode can usefully measure the parallaxes of bright stars
  at distances of up to $5 \kpc$, a factor of ten farther than achieved thus
  far with {\it HST}.  Long-period classical Cepheid variable stars in 
  the Milky Way, nearly all of which reside beyond $1 \kpc $, are especially
  compelling targets for parallax measurements from scanning, as they
  may be used to anchor a determination of the Hubble constant to $\sim 1\%$.
  We illustrate the method by measuring to high
  precision the parallax of a classical Cepheid, SY Aurigae, at a
  distance of more than $2 \kpc$, using 5 epochs of spatial-scan data
  obtained at intervals of 6 months.  Rapid spatial scans also enable
  photometric measurements of bright Milky Way Cepheids---which would
  otherwise saturate even in the shortest possible pointed
  observations---on the same flux scale as extragalactic Cepheids,
  which is a necessity for reducing a leading source of systematic
  error in the Hubble constant.  We demonstrate this capability with
  photometric measurements of SY Aur on the same system used for Cepheids 
  in Type Ia supernova host galaxies.  While the technique and results presented
  here are preliminary, an ongoing program with {\it HST} is collecting such
  parallax measurements for another 18 Cepheids to produce a better
  anchor for the distance scale.

\end{abstract} 


\section{Introduction} 

The increased precision of the cosmological model in the past decade
has been paralleled by a steady narrowing of the uncertainty in the
Hubble constant, the parameter which sets the present age and size
scale of the Universe. The first decade of the {\it Hubble Space Telescope
(HST)} was used to refine the value of the Hubble constant
(H$_0$) to $\sim \pm 10$\%, primarily by resolving Cepheids in distant
galaxies used to calibrate a diverse set of secondary distance
indicators \citep{freedman01,sandage06}. However, further progress from these 
distance ladders was restricted by systematic uncertainties in their rungs.  
\citet{riess11} sharply reduced the uncertainty to 3.5\% through four
improvements in the distance ladder comprised of Cepheids and Type Ia
supernovae (SNe~Ia): (1) calibrating eight modern SNe~Ia
with Cepheids, (2) observing Cepheids in the near-infrared (NIR) to
reduce the impact of extinction and metallicity, (3) the use of two
new geometric calibrations of Cepheids---parallaxes of Galactic
Cepheids from the {\it HST} Fine Guidance Sensor \citep[FGS;][]{benedict07}
and the 3\% geometric maser distance to NGC 4258
\citep[][and references therein]{humphreys13}, and (4) calibrating all
extragalactic Cepheid photometry with a single camera, WFC3, to negate
cross-instrument zeropoint errors. This measurement agrees within 2\%
with three subsequent determinations of H$_0$ \citep{freedman12, 
sorce12, suyu12b}, and it now constrains the
cosmological model with similar leverage as baryon acoustic
oscillations and high-redshift SNe~Ia when combined with the cosmic
microwave background data \citep[CMB;][]{sullivan11}.

Yet an outstanding enigma in the cosmological model remains.  What is
the nature of the dark energy accelerating the expansion of the Universe?  
A measurement of H$_0$ approaching percent-level precision would provide 
outstanding leverage for constraining the dark energy equation-of-state 
parameter $w = P/(\rho$ c$^2)$ \citep{hu05, suyu12a, weinberg13}.
Using the distance ladder to reach this goal requires an improvement in 
the measurement of its first rung, the nearby geometric determination of
distance.

Trigonometric parallaxes are the ``gold standard" of local distance
measurements---the simplest, the most direct, and the most
assumption-free.  They can be used to anchor an assortment of primary
and secondary distance indicators which sample the smooth expansion of
the Universe to measure the Hubble constant.

However, useful parallax measurements are among the most challenging
to acquire owing to the enormous distances to most objects of interest.
Only a few tens of systems have parallaxes greater than $0\farcs2$
($D < 5 \,\rm pc $).  Parallax measurements from ground-based facilities
have been limited by atmospheric refraction and flexure (gravitational
and thermal), with the zenith of efforts reached in the {\it General
Catalogue of Trigonometric Parallaxes} \citep{vanaltena95}, which contains 
parallaxes for 8,112 stars with a typical accuracy of $0\farcs 01$.  
{\it Hipparcos}, a space-based facility, measured parallaxes to a mean 
precision of $1 \mas$, providing parallaxes to 10\% precision or better 
for 20,853 stars, and to 20\% or better for 49,399 stars 
\citep{perryman09}.  More recently, the Very
Long Baseline Interferometer (VLBI) has been used at radio wavelengths
to measure parallaxes with $\sim 10 \muas$ precision beyond a kiloparsec for a
sample of sources distinguished by their radio brightness (primarily
star-forming regions, pulsars, and masers near asymptotic giant
branch stars) \citep{reid13}.  Unfortunately, this approach is not applicable to most kinds of stars.

Long-period Galactic Cepheids ($ P > 10 \,\rm days $) are among the most
coveted targets for parallax measurements, but they reside beyond the
capability of what has been feasible so far.  Cepheids remain such
vital targets because of their preeminent role as carriers of the
distance scale to extragalactic objects \citep{leavitt12}.  These
pulsating supergiants are relatively rare; several hundred are known
in the Galaxy, with distances $ D $ from the Sun ranging from 0.1 to
$ 10 \kpc $.  The closest, Polaris ($D=0.13 \kpc $), is less useful as a
distance indicator because it pulsates in an overtone mode. The
closest fundamental pulsator is the class namesake, $ \delta $ Cephei,
at $ D=0.27 \kpc $ and $ P = 5.37$ days .  Useful measurements beyond
$ 0.1 \kpc $ demand astrometric precision of better than $ 1 \mas $ and
repeatability over a year or more, a task made easier by the
resolution and stability of a space-based observatory.

\citet{benedict07} used 110 {\it HST} orbits and the two FGSs 
to measure parallaxes for 9 Cepheids with $ D < 0.5 \kpc
$ with an individual precision of $\sim 8\% $, providing one of the
best anchors of the measurement of the Hubble constant to date
\citep{riess09b, riess11, freedman12}.  (The FGSs provide the most
precise optical measurements of relative position, to 0.3
milli-arcseconds (henceforth \mas) for bright stars, a factor of 3 more
precise than the {\it Hipparcos} observatory).  However, the accuracy
of H$_0$ determined via the FGS sample is limited by several
factors: (i) its mean distance is determined to an accuracy of $ \sim
3\% $; (ii) they are too bright to image on the same {\it HST} system used to
observe extragalactic Cepheids along the distance ladder, introducing
an additional 2\% systematic error; (iii) their mean period ($\langle
P \rangle=6 \,\rm days $) is much shorter than those of the
extragalactic Cepheids visible at $D>20 \, \rm Mpc $ with {\it HST} in SN~Ia
hosts ($\langle P \rangle=30 \,\rm days $), propagating a systematic
uncertainty of 2\% in the Hubble constant per 5\% uncertainty in the
slope of the {\PL} relation. In addition, use of a simple linear
extrapolation of the Cepheid NIR {\PL} relation to bridge the period
gap between calibration and application is risky, especially given the
discovery of a break in the {\it optical} {\PL} relation at $P = 10$
days \citep{ngeow05b}.

Unfortunately, all long-period Milky Way (MW) Cepheids, except one,
are at $D > 1 \kpc $.  Conventional imaging limits the astrometric
precision of an unresolved, bright target to 0.01 pixel for WFC3-UVIS
or $400 \muas$ \citep{bellini11}, no better than a 2$\sigma$ detection
of parallax at $ 2 \kpc $.  To realize a 1\% calibration of the Hubble
constant, we need to calibrate a new sample of Galactic Cepheids whose
periods {\it bracket} those in extragalactic samples, while also
obtaining their photometry on the same {\it HST} system used for
extragalactic Cepheids.  Thus, we need to extend the range of {\it
HST}-based parallaxes to $D>1 \kpc $.  While the GAIA mission is
expected to obtain $ 10\hbox{--}20 \muas $ parallaxes across the sky
by the mission's end, the extreme brightness of long-period MW
Cepheids compromises its ability to accurately measure many of these
targets which have $ V < 7 $ mag; moreover, even with
successful parallax measurements from GAIA, the need for homogeneous
photometry of MW and extragalactic Cepheids will remain.

In pursuit of these goals, we have developed a new method for
obtaining useful measurements of stellar parallaxes with {\it HST} at
greater distances than previously possible.  {\it Spatial scans},
recently implemented on {\it HST} for WFC3-UVIS to measure exoplanet
transits \citep{mccullough11}, can be used to sample (i.e.,
reobserve) the target and astrometric reference stars in the field
thousands of times to provide an enormous boost in astrometric
precision over pointed imaging.  By scanning perpendicular to the long
axis of the parallax ellipse, this method can be used to improve the
precision of parallax measurements (and thus their useful range) by an
order of magnitude.  Spatial scanning also provides the means to
obtain reliable photometry of the MW Cepheids, whose exceptional
brightness would strongly saturate {\it HST} detectors in the briefest
possible exposures with conventional imaging.  

We have initiated three approved programs to collect an additional 18
sets of parallax measurements (GO 12879/13344 and 13334) and {\it
HST}-system photometry for these and other Cepheids (GO 13335).  Here we demonstrate
these techniques from a pilot program of spatial scanning observations
of the MW Cepheid SY Aurigae at an expected distance of about $ 2 \kpc $
and from calibration observations of the open cluster M35.  In \S 2
to \S 2.4 we describe the use of spatial scanning data to measure high
precision, relative astrometry at a single epoch.  \S 2.5 describes
algorithms used to combine multiple epochs of spatial scan data to
measure time-dependent astrometry.  \S 2.6 covers the use of the
preceding products to measure parallaxes.  In \S 3 we describe the use
of spatial scans to measure the photometry of bright sources.  In \S 4
we discuss future directions of spatial scanning observations.

\section{MW Cepheid Parallaxes with Optical Spatial Scanning}

%\section{MW Cepheid Parallaxes at $\rm D > 1 \kpc$ with Optical Spatial Scanning}

\subsection{HST as an Astrometric Platform}

%\subsection{HST as an astrometric platform; accuracy and the limitations of pointed observations}

Owing to its superior angular resolution and relative stability, {\it HST}
is a promising platform for obtaining high-accuracy relative
astrometry for sources within its field of view.  The theoretical
limiting precision for the measurement of a point source is approximately 
its root-mean square (RMS) width divided by the signal-to-noise ratio (S/N) 
of the observation.  The S/N in a normal exposure is
limited by the number of photons that can be collected before
saturation, typically $ \sim 10^5 $ for most imagers, thus suggesting
a theoretical limit of about $ 0.1 \mas $ per pointed observation.  In
practice, {\it HST} has so far been limited to a best-case
single-measurement precision of $\sim 0.3$--0.4 $\mas$ ($\sim 0.01$ 
WFC3-UVIS pixel; \citealt{bellini11}), both with the FGS
and with its imagers.  

There are several reasons for these
limitations.  With the most efficient imagers, light is discretely
sampled in pixels with angular size $\sim 40$ and 50 $\mas$ (WFC3
UVIS and ACS WFC, respectively).  Measuring the position of a source
to better than $\sim 1$\% of the pixel size has proved very
difficult \citep{bellini11}; among the contributors to
a noise floor may be zonal and temporal variations in the effective
point-spread function (PSF), and small-amplitude  irregularities in the geometric distortion which cannot reliably be
calibrated with existing data.   Any variations in the 
jitter over the image---produced, for example, by instantaneous rotations 
of the telescope---will introduce an additional field variation of the
effective PSF.  With the FGS, measurements are not limited by pixel
size, but they are carried out one star at a time, so stability of
the telescope and of the focal plane is paramount; in addition, the
limited rate at which photons can be collected with the phototube
detectors also limits the achievable S/N
independent of the source brightness.

%\subsection{Astrometry with optical scans; increased sampling and jitter removal}

\subsection{Astrometry with Optical Scans}

Many of the limitations of pointed observations can be overcome via a
new observing mode with WFC3, spatial scanning under FGS control,
developed for WFC3 in 2011 to obtain photometry during exoplanet
transits.  In this mode, the target field is observed while the
telescope is slewing in a user-defined direction and rate (to a
maximum of $7\farcs8$ s$^{-1}$).  Up to $5\arcsec$ s$^{-1}$, the
telescope can maintain FGS guiding at all times; for faster scan rates,
the telescope must be controlled by gyroscopes, leading to a smoother but
less accurate motion.  Each source thus describes a ``trail'' on the
detector (see Figure 1).  In the simplest mode, the motion is straight and uniform,
resulting in a straight trail with constant brightness (counts per
unit length) after accounting for geometric distortion in the
detector.  The trails for all sources are parallel in the
distortion-corrected frame.  More complex ``serpentine'' scans are also
possible and can be preferable for sparse fields or exceptionally
bright targets (see \S4).

Some of the advantages of this method are immediately obvious.  The
light from each source is spread over a much larger number of pixels,
allowing a larger global S/N to be achieved for each
source\footnote{The original motivation for this mode was the ability
to collect $> 10^8$ photons per source without saturation, thus
allowing high-precision global and time-resolved photometry of bright
sources such as stars with transiting exoplanets.}.  Furthermore, each
long trail provides thousands of separate position measurements in the
cross-trail direction, one for each pixel traversed, thus averaging
out the impact of single-pixel and local irregularities.  Scanning at an angle relative to the detector also provides
sub-pixel sampling of the undersampled WFC3 PSF. 
Because the
measurements are time-resolved (e.g., 25 pixels per second at a scan
rate of $1\arcsec$ s$^{-1}$), the telescope jitter can be subtracted as a
function of time, negating the impact of even large ($\sim 1$ pixel) 
jitter events which are not uncommon.  With spatial scans, we
expect to routinely achieve measurement precision of one-thousandth of
a WFC3-UVIS pixel (1 millipixel, or {\mpix}, corresponding to $40 \muas$) 
or better.

The disadvantage of this method is that precise measurements can be made 
in only one direction at a time, the direction perpendicular to the scanning
motion, as positions in the direction along the motion are blurred by
the motion itself.  Thus, a precise two-dimensional measurement of
relative positions requires in principle two observations.  It is
advantageous to choose scans to occur along the parallel readout
direction (i.e., the $Y$-detector axis), as this limits the dominant
direction of imperfect charge transfer and its attendant smearing to
occur along the direction not being measured.  The much smaller effect of the
{\it serial} charge transfer efficiency is addressed in \S 2.5.2.

The disadvantage of obtaining positional changes in just one dimension
is minimal for the measurement of parallaxes, as the motion of interest
takes place in a predictable direction.  By choosing the scan
direction appropriately, the measurement can be made for the
dominant parallax component.

\subsection{Designing the Observation}

% source S/N, source catalogue

The main characteristics of the planned observations relate to the
brightness of the source, the availability of reference stars within
the detector field of view, and the desired timing of the observation
vs.~the allowed telescope roll angles.  An unusually large degree of 
planning and simulating is needed to obtain useful observations in this mode.

As in all narrow-field astrometric observations, WFC3 spatial scan
observations can only measure the {\it relative} parallax of the
target---the difference between the parallax of the target and
that of nearby reference stars.  All stars in the field of view move
along similar parallactic ellipses with the same shape, orientation,
and phase because the apparent parallactic ellipse traced annually on
the sky is simply the reflex of the motion of the Earth around the
barycenter of the Solar System.  However, the amplitude of the motion
of each star (e.g., its semimajor axis) scales inversely with its
distance from the Sun, and represents the parallax.  Since the
absolute pointing of each observation is not known to better than a
few tenths of an arcsecond, only the {\it difference} between parallaxes
of stars in the field can be measured to useful precision.
\footnote{This situation is different from that of a telescope with
wide-angle capability, such as GAIA or {\it Hipparcos}, which can observe
simultaneously stars in different regions of the sky separated by a
``basic angle'' of tens of degrees, which then follow parallactic
ellipses having different shapes and at different phases.  Absolute
parallaxes can be determined with precise knowledge of the basic angle
and its stability, but they can be sensitive to systematic uncertainties 
in these quantities.  Thus, parallaxes measured from the wide and
narrow-field approaches provide a useful test for systematic errors in
either one.}

For this reason, it is critical for the field to contain a sufficient
number of reference stars with independent distance estimates to
correct relative parallaxes to absolute.  Our typical observations in
the plane of the Galaxy measure between 50 and 200 stars in a field of 
$5\farcm \times 2\farcm7$ defined by scanning WFC3-UVIS.  It is important to
note that distant stars, such as K giants several magnitudes fainter
than the target Cepheid, are especially valuable in providing a
correction to absolute parallax.  For example, if a reference star is
at $ 5 \kpc $ (thus a parallax of $ 200 \muas $), even a relatively crude
estimate of its absolute magnitude, perhaps with a 0.3 mag error, 
leads to a $ 30 \muas $ uncertainty in its contribution to the correction to
absolute parallax.  With several such stars (a typical field will
have 2--10 at this distance or beyond) and the use of a
comprehensive set of UV, Str\"{o}mgren, broad, and NIR bands to estimate
stellar parameters, the uncertainty in this correction is expected to
be well below our target measurement precision of $ 1 \mpix $ ($ 40
\muas $).  Nonetheless, the correction to absolute parallax is a
critical feature of our program, and the related uncertainties will be
discussed in \S 2.6.

The optimal duty cycle for observing utilizes an exposure
time of 350 s, just long enough to dump the buffer containing
the previous observation during the next exposure, and a scan length
of $ \sim 144 \arcsec $, as long as possible while ensuring that both
the start and end points of the scan are visible for the primary
target within the $ 163 \arcsec $ field of view.  This combination
allows the largest number of scans per orbit, 4 or 5 depending on
visibility, and near-continuous observations, and it yields a fiducial
scan rate of $ \sim 0 \farcs 4$ s$^{-1}$.

While any filter may be used for scanning, it is best to use a filter
which has low wavefront errors and that passes light where CCD
fringing is low.  The use of longer wavelengths and redder filters would degrade
resolution and increase fringing.  Working too far to the blue can
severely undersample the PSF (by a factor of 2 at 4400 \AA) and
diminish the S/N of red giants, which provide the best calibration to
absolute parallax.  The throughput of the filter is an important
consideration as it determines the S/N of the target within the
limited, useful dynamic range ($6 \times 10^4$ $e^-$ pixel$^{-1}$ to
$3 \times 10^2$ $e^-$ pixel$^{-1}$, about 5.5 mag).  It is also
advantageous to choose a filter with a well-calibrated geometric
distortion field \citep{bellini11}, as this provides the source of
initial transformation from detector coordinates to the sky.  The
{\Filter{F606W}} filter ($\Delta \lambda = 2300$ \AA, $\lambda_0 =
5907$ \AA ) is an attractive choice which makes useful measurements at
the fiducial scan rate for stars at $10.6 < V < 16$ mag.  Stars at
$D=2 \kpc$ in this brightness range will have $-1 < M_V < 4.5$ mag.
Red giants in this brightness range will be at $5 < D < 10 \kpc $ (all
magnitudes are quoted in the Vega system).

Long-period Cepheids ($ P > 10 $ days, $ -4 > M_V > -7 $) within $ 3 \kpc $ 
are more challenging parallax targets because they are very bright ($6 < V 
< 10$ mag) and thus would saturate in a broad-band exposure at the fiducial 
scan rate.  In order to avoid saturation, either a faster scan or a narrower
filter is needed, with the reduction in counts proportional to the
inverse of the increased speed or the decreased filter width.  Scans faster than $5\arcsec$ s$^{-1}$ 
rely on less precise gyro guiding, and given the importance of
maintaining a fixed scan direction, we prefer to avoid their use. 
In most cases, scans at a similar rate but employing a medium- or narrow-band 
filter---e.g., {\Filter{F621M}} ($\Delta \lambda = 631$ \AA, saturation
limit $V=9.2$ mag) or {\Filter{F673N}} ($\Delta \lambda = 100$ \AA, 
saturation limit $V=7.2$ mag)---suffice to observe the Cepheids without 
saturation.\footnote{Accounting for the variation in the Cepheid brightness 
can be important in fine-tuning the saturation limit.  Such fine-tuning 
requires knowledge of the Cepheid phase and use of highly constrained 
observing times.  It is feasible to observe at $V<7$ without saturation using faster scans and serpentine scanning
to make effective use of the additional scan length available.}

However, in such observations there are typically too few (5--10)
sufficiently bright stars that can be measured with enough precision
to provide the desired target precision in the reduction to absolute
parallax (based on the shot noise of the source, $ 20 \muas $
precision requires $ \sim 1000 $ counts per row and a trail length of
a few thousand pixels).  Thus, we rely on a hybrid approach: shallow,
narrow-band scans to measure the Cepheid and a few reference stars,
followed by deep, broad-band scans to measure 50--200 distant
reference stars.  The stars bright enough to be measured in the
shallow scan and yet faint enough to avoid saturation in the deep scan
serve as a bridge between the target Cepheid and the bulk of the
reference stars; for this purpose, we do not require any knowledge of
their distance or motion.  Since the broad and shallow scans are taken
concurrently, there is no significant astrophysical motion of the
reference stars between scans, and the uncertainties associated with
long-term motions (e.g., parallax) are negligible in this step.
However, linking together shallow and deep scans can be a significant
component in the final parallax error budget, as discussed in \S 4.

\subsection{Analysis of Scan Data}

Figure~1 shows a typical scan image for the SY Aur field. (See Table~1
for a listing of scanning observations of SY Aur.)  The nearly vertical
``trails'' are the images that each star leaves as the telescope scans
over $ 144\arcsec $, 88\% of the length of the field of view.  The
area covered by the scan is almost twice the normal field of view of
the camera.  Stars near the center of the region spanned in the
detector $ Y $ direction will have trails that start and end within
the frame, while stars farther from the center have trails that enter
or leave the frame during the scan.  Cosmic rays are the only compact
sources in the frame and are readily identified by their lack of
vertical extent, allowing us to disregard impacted pixels.

\subsubsection {The M35 Calibration Program}

As part of the WFC3 calibration program, multiple scan observations
were obtained in {\it HST} program 13101 for a field covering part of
the open cluster M35 at two orientations $ 180\degr $ apart, using the
filters {\Filter{F606W}}, {\Filter{F621M}}, and {\Filter{F673N}}.  See
Table~1 for a listing of these observations.  Owing to its position
very close to the ecliptic plane, M35 can be observed at nearly
constant telescope roll angle for several months, and it reverses
available roll within a few days of its antisun position.  It also
includes 40 stars of nearly optimal brightness ($10<V<17$ mag) for
observations in the selected filters.  This combination of properties
makes it an ideal target for calibrating spatial scan astrometry and
in particular parallax observations, which also require observations
at orientations differing by $ 180 \degr $.  The stars in M35 have a
small velocity dispersion of $ 0\farcs02$ per century ($ 0.5 \muas$
per day), so the few-week interval between observations at
orientations that differed by $ 180 \degr $ produces $\sim 0.25 \mpix
$ of dispersion; thus, the relative positions of the stars can be
treated as being static within the calibration data.  At each of two
epochs a sequence of 350 s scanning frames of the M35 field were
obtained back-to-back, 10 for {\Filter{F606W}} and 5 each for
{\Filter{F673N}} and {\Filter{F621M}}, to better understand the effect
of the orbital thermal cycle and $ 180 \degr $ orientation change on
relative positions.  We will refer to a number of results derived from
the analysis of the M35 data in the following sections.

\subsubsection {Matching Trails to Stars}

The first step in the analysis of scan data is the identification of
the trail positions and of the stars associated with each trail.
Using a catalog of star positions obtained from direct WFC3-UVIS
images, the start and end positions of star trails can be predicted
from the scan location and its velocity vector.  Small adjustments are
needed to precisely align the direct and scan images, whether both are
obtained at the same epoch or not.

It is also necessary to identify and catalogue stars with trails that
overlap trails to be measured.  We refer to these contaminating
sources as ``spoilers,'' as they can throw off subsequent position
measurements.  We identify the regions within each star's trail that
are affected by spoilers, and disregard the impacted pixels.  Tests show
that spoilers are significant if the stars differ by $< 7$
mag and the trails lie within 8 pixels of each other.

\subsubsection {Minirow Fits}

The next step in the analysis process is measuring the cross-scan
position of each star as a function of position along the trail.  For
each trail, we extract the pixel values and quality flags at each
integer pixel step along the trail within $ \pm 10$ pixels ($\pm 4
\times$ the full width at half-maximum intensity) of the nominal trail
center.  We call this individual block of information a ``minirow,''
as it is a small part of the detector row where signal from the star
of interest is located (see Figure~2).  Individual trails may be
composed of a few hundred to several thousand minirows.  \footnote{We
discard trails with fewer than 300 available minirows, as their
astrometric fit will generally suffer in both statistical value and
systematics from the small coverage.}  Each minirow thus consists of a
short (typically 21 pixel) array of data values vs. $ X $ pixel
location at a fixed $ Y $ location.  Each pixel is also assigned a
weight based on the detector noise model including zero weight for bad
pixels; pixels too close to a spoiler are given zero weight, and if
pixels within $ \pm 2$ pixels of the peak are rejected, the minirow is
rejected as being invalid.  The fit involves three parameters: the
amplitude, or scaling factor for the line-spread function (LSF), which
sums to unity over its full length; the center position, or amount by
which the LSF must be offset and spline-interpolated along the
detector $ X $ direction for an optimal match; and the background, or
constant level that must be added to the line profile to match the
data.  The LSF, oversampled by a factor of 4, is previously derived as
a function of $ X $ and $ Y $ detector position from the empirical PSF
in images of star clusters \citep{bellini11}.  Positional
uncertainties are determined from the $\chi^2$ of the minirow fit with
a minimum floor of 0.01 pixel imposed at $10^{4}$ $e^-$ pixel$^{-1}$
to reflect the finite precision of the geometric distortion field
obtained from \citep{bellini11}.

\subsubsection {Position in Rectified Coordinates}

For each trail, the minirow fits yield a number of $ (Y, X) $ pairs
representing the fitted $ X $ position in detector space at the $ Y $ detector
location.  In order to proceed with the analysis, we need to
transform these $ (Y, X) $ positions into a reference frame in which they
are directly comparable; for example, we expect the trails from
different stars to be essentially parallel, with their perpendicular
(across scan) separation constant; this is not true in detector
coordinates, which suffer from a variable geometric distortion.  In
addition, portions of the trails in Detector 1 cannot be readily
compared to those in Detector 2 without a global distortion solution.

We use the field distortion solution from \cite{bellini11}, which uses
a definition of the PSF position that is consistent with the empirical
determination of the PSF.  Whatever the definition of the PSF position
(e.g., centroid, center of symmetry, peak), it is crucial that it
coincide with the derivation of the field geometric distortion.  
{\it From here forward our analysis utilizes pixel positions on the 
sky unless explicitly stated otherwise.}

The geometric distortion solutions have an accuracy of
$\sim 0.01$ pixel on scales of $\sim 40 $ pixels
\citep{bellini11}.  This accuracy is
sufficient to reach position precision of 1 {\mpix} ($ 40 \muas $)
for full-length scans, which would be a significant contribution to
our overall error budget.  In addition, the thermal cycle of {\it HST}
within an orbit (also known as telescope ``breathing'') and temperature variations
due to the recent attitude of the telescope
changes the
focus position by $ \pm 5 \micron $ during the orbit, which can cause
the optimal geometric distortion solution to vary.  Although a thermal
model of {\it HST} is available to predict these variations, the
optical model of the telescope does not have enough fidelity to allow
a useful correction of the geometric distortion field.  An improved
geometric solution which takes into account its possible variability
is necessary to achieve our measurement goal.  In the following
sections we show how a combination of internal and self-calibration
can be used to reduce the impact of the geometric distortion below $
\sim 20 \muas $.

Annual variations in Earth's velocity vector induce changes in plate scale
which can be treated as a simple isotropic scale term for our narrow
field.  Typical values of the velocity aberration are of order $
10^{-4} $ from the $\pm 30$ km s$^{-1}$ Earth velocity, requiring a
scale correction of $ \pm 200 {\mpix} $ from the frame center to the
edge.  Variations in spacecraft velocity during a scan result in
variations in the velocity aberration which can reach a peak of $ 6
\times 10^{-6} $ or $ \pm 10 {\mpix} $. This velocity aberration correction,
proportional to the combined velocity of the telescope and Earth with
respect to the Solar system barycenter, projected along the line of
sight to the target, is computed on the basis of the {\it HST}
ephemeris as part of the standard pipeline image processing, and is
stored in the header of each image.  We apply a simple sinusoidal
interpolation to the recorded values in order to obtain the
{\it instantaneous} scale correction for each minirow---the value depending
on the time attached to that minirow, or equivalently, the time at
which the star was traversing that specific pixel location.  All
position measurements are scaled according to the instantaneous
velocity aberration that applies.

\subsubsection {Jitter and the Reference Scan Line}

A feature that is immediately apparent from the values of the fitted 
$X$ position vs.~position along the scan (see Fig.~2) is that the
values vary locally much more than the estimated error in each
measurement, with variations of $\sim$ 0.1 pixel and
occasional jumps of up to 1 pixel.  Comparing different trails at
equivalent positions along the scan (e.g., at the same distance from
the start of the scan), as shown in Figure~3, reveals that these
variations are highly correlated from star to star, and are in fact
caused by irregularities in the telescope motion---equivalent to the
so-called ``jitter'' in pointed observations.  Unlike pointed
observations, for which the jitter produces
a modest blurring of the PSF, scanned observations can be used to
measure and correct for the telescope jitter perpendicular to the
direction of motion (our preferred measurement axis).  In addition,
the presence of jitter allows a more accurate measurement of the
relative position of each star {\it along} the scan direction, from a
template fit or cross-correlation of the jitter features.  Each star
will be affected by different jitter features, depending on where it
is located in the field of view and thus for what fraction of the
exposure it is on the detector; to avoid systematic offsets due to
asymmetric jitter, it is thus necessary to correct for the jitter
at the pixel-by-pixel level.

For this purpose, we define a {\it reference scan line} from the
weighted average of all stars.  The reference scan line represents the
instantaneous offset from the ideal rectilinear motion of the
telescope as a function of position along the scan (which is a proxy
for time).  A proper definition of the reference scan line requires
accurate ($ < 0.25 $ pixel) knowledge of each star's position along
the scan direction, to avoid blurring of jitter features.  We carry
out this process iteratively.  We obtain a first approximation to the
reference scan line from the along-scan position determined either
from a star catalog or from the apparent start/end position of the
trail; these are typically accurate to $ \sim 1 $ pixel.  Then we fit
this approximate reference scan line to each trail, thus obtaining a
better position along the scan, and produce a new reference scan line
from the updated positions in rectified pixel space.  The final
reference scan line is oversampled by a factor of 4; for the
best-measured trails, the estimated uncertainty in the along-scan
position is a few percent of a pixel, sufficient to resolve the jitter
frequency.  Figure~3 shows the comparison of bright star trails
aligned in scan time and the residuals after subtraction of one from
another.

\subsubsection {Trail-to-Trail Separation Along the Scan: Variable Rotation}

One fundamental assumption of the scan method is that trails for
different stars are essentially parallel, so their separation
remains constant (on the sky, after correcting for geometric distortion)
throughout the observation.  On short time scales ($ < 1 $ second),
the parallelism of trails is demonstrated by the correspondence of
their jitter features.  However, on longer time scales ($ \gtrsim 1 $
minute, corresponding to several hundred pixels traversed), the
separation of trails occasionally appears to vary by several
{\mpix}---difficult to measure individually, but quite apparent in a
statistical sense.  Initially we investigated the possibility of
deviations from the nominal geometric distortion; however, this
interpretation would require that the deviations be repeatable across
observations of different fields, which turns out to be inconsistent
with the properties of the calibration data obtained for M35.
Instead, the nonparallelism of trails exhibits a pattern consistent
with slow rotations of the telescope's field of view (equivalently,
its roll angle) throughout the observation.  Under this
interpretation, a single parameter---the instantaneous differential
roll angle---causes small changes in the separation of each pair of
trails in a pattern strictly dependent on their relative separation in
the scan direction.  Figure~4 shows the scan patterns formed by fixed
or variable field rotation.  Indeed, solving for a slow variation of
the telescope roll angle (typically fitted as a fifth-degree
polynomial in position along the scan, a proxy for time) removes the
variation in separation between trails below statistical significance.
Empirically, the instantaneous roll angle changes by $ 0.001 \degr $ to $
0.003 \degr $ over an observation.

The most likely reason for a variation in the instantaneous roll angle
is in imperfections of the geometric solution in the FGS.  Under FGS
control, the telescope tracks the position of the guide stars
throughout the observation, and makes continuous roll adjustments on
the basis of their position in the focal plane to ensure that the roll
angle remains constant.  However, this requires very good knowledge of
the geometric distortion solution in both FGSs over the more than $ 2\arcmin
$ that each guide star traverses during a scan.  A local error of $ \sim
0.5 \mas $ over $ 30\arcsec $ would suffice to explain the observed
changes in roll angle.  Crucially, we observe that the variation in
roll angle is consistent across observations of the M35 field,
regardless of filter, direction, and time of the scan; however,
different fields, or observations of the same field at different roll
angles (which changes the position of the guide stars within the
FGSs), have essentially a different but again internally consistent set of
variations of roll angle.  Both of these properties are consistent with
the FGS geometric distortion interpretation \citep{nelan12}.

In practice, we determine the change in roll angle as a function of
position along a scan field using an average obtained from deepest
scans of that field.  The mean is then applied to all scans of this
field.  This is especially helpful for shallow scans whose trails lack
the precision necessary to effectively quantify the variations in roll
during the scan.

The result of the analysis of each scan image {\it treated
individually} is a set of position measurements for each star in the
nominal $ X $ direction.  These measurements are with respect to the
reference scan line defined after the correction for variable roll.  The position of the reference scan line is
arbitrary; there is information only in the {\it relative} position of
the stars with respect to one another.\footnote{The absolute
pointing precision of {\it HST} with respect to an inertial reference system
is several orders of magnitude less accurate than the precision goal
of this project.}  Each measurement is also associated with its
estimated error, and several ancillary quantities are recorded for each
trail, including the number of valid minirows, typical amplitude and $
\chi^2 $, and the position of the trail on the detector.

By measuring stellar positions at three or more epochs at six month intervals,
we may now determine the proper motion and parallax of the
measured stars.  

% Section III.B: The single-epoch aggregation step

\subsection {Using Multiple Observations at the Same Epoch}

For some applications of astrometric measurements it may be beneficial
or even necessary to first combine the measurements of multiple scans
obtained at a common epoch.  Applications include combining multiple
scans to reduce sources of error, combining scans obtained under
different conditions to calibrate the effect of the conditions on
measured positions (e.g., thermal state), or combining scans of
differing depth to improve the dynamic range of the measurements.  In
applications for which the target and reference stars can be well
measured at a single exposure depth (i.e., a range of $\sim 5$
mag), one can proceed directly to measuring parallax in \S 2.6.

\subsubsection {Bridging the Dynamic Range Gap for Cepheids}

Because of the large difference in brightness between Cepheids and
field stars, typically 4--10 mag, it may be necessary to obtain shallow
and deep exposures at the same epoch and combine their measurements.
To obtain high-quality measurements at the fiducial scan speed for
reference stars in the range $11<V<17$ mag, a depth where they are
plentiful, {\Filter{F606W}} is the filter of choice, with a typical field
providing 50--200 such stars with a full-scan uncertainty $ < 50 
\muas $, sufficient to control the uncertainty associated with the
reduction to absolute parallax addressed in the next section.  A
narrower filter---e.g., {\Filter{F673N}} or {\Filter{F621M}}---is 
then used to obtain an
unsaturated trail for a bright Cepheid with $ V < 10 $ mag; however, in
shallower scans most reference stars are too faint to provide a good
anchor to spectroscopic distance estimates and thus a reduction to
absolute parallax.  

Our solution for SY Aur ($8.6<V<9.4$ mag) is to obtain a pair of
scanning observations in both {\Filter{F606W}} and {\Filter{F673N}}
within the same orbit, during which the stars can be assumed to be
stationary.  The shallow observations in the narrow-band filter yield
a full-precision measurement for the Cepheid (indeed, we may choose the
filter and scan speed so as to maximize the electrons collected from
the Cepheid while avoiding full-well saturation) and good measurements
for $ \sim 5$--10 other stars of intermediate brightness in the field.
The intermediate stars anchor the shallow frame to the deep frame, and
thus tie the Cepheid motion to the reference frame defined by the
collection of reference stars.  We call this step ``aggregation''; it
results in position measurements for all stars, including the Cepheid,
in a consistent epoch-based reference frame.  Errors are also reliably
estimated from the aggregation step, with contributions from the scan
measurements and nuisance parameters used to combine scans, and the
resulting measurements at each epoch form the basis for the parallax
and proper-motion measurements described in the following section.
Note that a dearth of intermediate brightness stars can lead to
degradation of the Cepheid parallax, as discussed in section \S 3.

\subsubsection {Single-Epoch Position Measurements}

The goal of the aggregation step is to obtain for each star a single
measurement of position perpendicular to the scan direction that
optimally uses all the measurements obtained at the same epoch.

The aggregation step uses a model relating the ``true'' position of each
star in the measurement direction to its measured position in each
scan.  Each star has one free parameter, its nominal position in the
measurement direction; in total, there is one fewer parameter than
stars, as our position measurements are {\it relative}, and thus
insensitive to a bulk shift of all positions in the measurement
direction.  As a convention, we take the mean of the true positions of
all stars observed in the field to be zero to set the arbitrary
reference position.  The true positions of $ N $ stars are thus
described by $ N-1 $ positional parameters.

In addition to the parameters describing the stars' true positions,
the transformation from true to observed position involves several
additional effects that need to be quantified and modeled.  The most
obvious ones are scan-to-scan offset and field rotation as discussed in \S 2.4.5 and 2.4.6.  The 
offset is a free parameter for each scan, as the absolute telescope
pointing is not stable enough to constrain its scan-to-scan position
to the required accuracy of tens of $ \muas $.  The roll angle of
the telescope is not identical for scans with the same requested roll angle,
and small variations ($ <
0.001\degr $) need to be taken into account.  Both offset and rotation
are defined with respect to the first scan by convention; thus,
aggregating $ M $ scans requires $ M-1 $ offsets and $ M-1 $
rotations.

Analysis of the M35 data, seen in Figure~5, shows clearly that
these two parameters are not sufficient.  Residuals between model and
measurements routinely reach 5--10 $\mpix$, even for scans that
are obtained in the same filter and orbit in succession, well in
excess of the expected statistical measurement errors.  Furthermore,
such residuals, as shown in Figure~5, have clear spatial correlations
that indicate that there is variable low-order distortion at the few
{\mpix} level.  The residuals are somewhat greater between different
filters, largely due to a static scale difference, and between
observations obtained at different orientations, as these include both
the previous variable term as well as errors in the static geometric
distortion field.  These residuals do not show any correlation with
the sources' brightness or color, and thus are most likely related to
variations in the geometric transformation between true and measured
positions.

For the M35 calibration data, a successful approach consists of
including in the model a low-order polynomial correction with free
coefficients for each observation (see Fig.~5).  We adopt a
polynomial correction to the $X$ coordinate when aligning two frames
which depends on the pixel position of each star trail in the
detector; the assumption is that any variation in the transformation
from true to measured position is tied to the telescope and detector,
and therefore is best described in measured rather than true
coordinates.  Note that a generic first-degree polynomial includes by
definition an $ X $ scale term (the first-order correction in $ X $)
as well as a {\it detector} rotation, which is slightly different but
closely related to the {\it field} rotation previously considered.

We find that a second-degree polynomial as a function of $ X $ and $ Y
$ coordinates is adequate to describe the $ X $ coordinate
transformation between two scans and reduce residuals of well-measured
stars to $ \lesssim 1 \mpix $, as shown in Figure~5. (Only terms of
{\it total} degree up to the polynomial degree are included.)
Figure~5 shows the residuals after these polynomial corrections are
included in the model.  A second-degree polynomial in $ X $ and $ Y $
has five coefficients and the transformation has a total of seven
parameters, including rotation on the sky and a constant term.  The
leading polynomial term of the cross-filter match has degree (1,0),
corresponding to an overall scale factor, and indicates that the
geometric solution we initially adopt has a scale in {\Filter{F673N}}
that is about $ 1.5 \times 10^{-5}$ different from the
{\Filter{F606W}} scale, or up to $40 \mpix$ near the edge.  Several
other terms also have significant power but generally vary in sign,
with contributions of up to $ 50 \mpix $ at the edge of the field.
Not surprisingly, the coefficients of these terms are smallest when a
pair of frames have a similar predicted time-dependent focus position
of the telescope based on the {\it HST} orbital thermal model as shown
in Figure~5.  Thus, it appears that much of the
role of the polynomial correction is to account for PSF and field
distortion caused by the changing thermal state of {\it HST}.

To quantify the size of a color dependence of the geometric distortion
(i.e., a color wedge) over the broad range of the {\Filter{F606W}}
filter, we used the M35 data to compare the relative source positions
in {\Filter{F673N}} and {\Filter{F606W}} versus the $ B - V $ color
after accounting for variable distortion.  The full color range of the
stars was $ -0.15 < B-V < 1.4 $ mag, with a mean color of 0.57 mag.
We assume that any color effect in {\Filter{F673N}} is negligible
owing to its very narrow wavelength range ($\Delta \lambda= 100 $ \AA\
vs $\Delta \lambda= 2300 $ \AA), so the dependence on color is a
measure of a chromatic term in {\Filter{F606W}}.  We find a very small
color shift from the M35 data of $ 0.6 \pm 0.2 \mpix $ for an 0.5 mag
difference in color from the mean.  Eliminating the two bluest stars
with $B-V<0$ mag reduces the effect further to $ -0.1 \pm 0.2 \mpix $
and is more appropriate for the colors encountered for field stars and
Cepheids.  Given the small empirical size of a chromatic shift and the
fact that Cepheids lie near the middle of the color range of stars, we
conclude that a chromatic effect in {\Filter{F606W}} is negligible,
and we will explore additional chromatic effects with an expanded
dataset in the future.

In principle, imperfect charge-transfer efficiency (CTE) can smear
charge along the readout axes and shift astrometry.  The effect is
1--2 orders of magnitude smaller along the $X$ axis for {\it HST} CCDs
owing to the greatly reduced pixel transfer (and trapping or
detrapping) time than for the $Y$ axis.  For this reason we chose to
scan observations along the Y-axis and measure astrometry along the
X-axis.  However, a small amount of charge is apparent trailing hot
pixels in the $X$ axis opposite the read direction indicating the
presence of short timescale charge traps and imperfect X-CTE.  This
deferred charge is about 0.08\% for a pixel with 10,000 electrons and
becomes a smaller fraction for brighter pixels.  The resulting
astrometric shift is about twice as large for a given charge total in
a 2 dimensional PSF than the 1 dimensional minirow of a scan due to
the larger shift affecting the fainter PSF wings.  Using these
calibrations, we estimate the shift of SY Aur in the shallow scan from
fast trapping would be $\sim 0.4 \mpix $ and even smaller, about $<
0.2 \mpix $, for a Cepheid with peak counts of $> 30,000 $ electrons.
Depending on the relative position of stars and amplifiers across
epochs, the effect on a star's measured parallax may fully or
partially cancel.  For SY Aur we conclude that the effect ($< 3\% $ at
$ 2 \kpc $) can be neglected.  In the future we will try to directly
measure the change in scan line astrometry in the X-direction by
dithering a source back and forth across the line dividing the
read-out X-direction for use in a larger sample analysis.

For the SY Aur field we observed the Cepheid at $ 8.7 < V < 9.4 $ mag
in {\Filter{F673N}} at the fiducial scan rate of $ 0\farcs 4 \,\rm
s^{-1}$, resulting in peak counts for the Cepheid of 10,000--15,000
electrons, thus providing unsaturated measurements to complement the
contemporaneous {\Filter{F606W}} scans (where the Cepheid is
oversaturated by a factor of 5--6 and {\Filter{F621M}} would saturate
by a factor of 1--1.5 if used).  Unfortunately, the
anti-galactocentric direction of SY Aur (Galactic longitude $ \ell =
164.75\degr $) results in a rather sparse field with only three stars
that are both well-measured in {\Filter{F673N}} ($7 < V < 13$ mag) and
not saturated in {\Filter{F606W}} ($ V > 11$ mag).  Because the
calibration requirements to reach millipixel precision from spatial
scanning were not known at the start of this pilot project, the
selected field of SY Aur is less than optimal.

A few more stars exist at $ 13 < V < 15 $ mag, providing a little
additional constraining power.  The paucity of stars available to
define the transformation between the shallow and deep frame results in
a relatively noisy mapping which is not very robust.  While the mean
precision of each Cepheid position measurement in {\Filter{F673N}} is $ 0.4
\mpix $ (see Fig.~6), transforming the unsaturated Cepheid position
in {\Filter{F673N}} to {\Filter{F606W}} with so few stars degrades the Cepheid
position to about $ 2 \mpix $ precision.  As shown in Figure~6, for
other Cepheid fields selected in directions richer with reference
stars, we find that we can constrain the Cepheid position in the deeper
scan to $ 1 \mpix $.

\subsection {Multiepoch Combination and Parallax Fit}

The final step is to utilize the multiple measurement epochs taken
over the course of two years at intervals of six months, in order to
estimate the parallax and proper motion of each star in the field.
This model involves several considerations.  First, the position of
the stars in the field, relative to one another, naturally changes
over time as a consequence of their astrometric motion {\it in the
measurement direction}.  This motion is modeled as the combination of
a linear term, which is the projection of the proper motion along the
measurement direction, and a periodic term, which is the projection of
the parallactic motion.  Note that the shape and phase of the
parallactic motion as well as the component in the measured direction
are fully known from the position of each star, the motion of the Earth
with respect to the barycenter of the Solar System, and the
orientation of WFC3; the only free parameter is its amplitude, which
scales directly with the parallax of the target.  Thus, the standard
astrometric model involves three parameters for each star: position,
parallax, and proper motion along the measurement direction.

Observations spaced every half year, at the time of the maximum and
minimum parallactic excursion, give optimal discrimination between
parallactic and proper motion. In practice, {\it HST} observations of 
SY Aur could be scheduled at the optimal time (mid September and
early March) at an orientation no closer than $ 35 \degr $ (or $
180+35 \degr $) from the optimal, reducing the apparent parallax
component to 80\% of its maximum value. The available roll angle at a
given time depends on {\it HST}'s orientation with respect to the Sun
(in relation to the telescope aperture and solar panels) and the
availability of guide stars.  Although the observation orientation may
not be optimal, symmetry ensures that the same orientation flipped by $ 180
\degr $ will be available 6 months later.

At least three epochs are necessary to be able to disentangle parallax
and proper motion for each star.  In practice, three epochs will not
suffice; additional free parameters are involved in
registering the observations, and the errors in the derived
parameters would be much larger than the measurement errors.  Four
epochs are in general sufficient to obtain good constraints on the
parallax and proper motion separately; when available, five epochs
help reduce the covariance between derived parallaxes and proper
motions, and improve the precision of the final measurement beyond the
obvious factor $ \sqrt {5/4} $.  For SY Aur we have in fact five
epochs, although two lack repetition of the shallow or deep scan.

In addition to the astrometric parameters of each star, the model
includes geometric parameters used to align each epoch with one
another (offset and rotation), as well as any residual large-scale
adjustment to the geometric distortion required to reduce the model
residuals.  This last part is identical to the single-epoch aggregation 
step in \S 2.5.2, but it now substitutes the stationary star assumption
with the astrometric model for each star.  We also now utilize epochs 
obtained with orientation differences of 180$^\circ$.

The full model can be formally described by the expression

\bq X_{ij} = X_{i0}  - X_{\rm ref,j} + pmx_i \, (t_j-t_0) + \pi_i \, f_j +  R_j\, Y_{i0} + \langle P(X_{\rm det}, Y_{\rm det}) \rangle_{\rm trail,ij} \eq

\noindent
Here the basic measurements are the positions $ X_{ij} $---that is,
the $X$ position of the trail of star $ i $ in image $ j $ (relative
to the reference scan line), measured after correction for variable
rotation, scale-corrected for velocity aberration and variable
distortion, and projected onto a constant sky frame.  The $ X $
coordinate is aligned with detector $ X $ and, by design, aligned with
the bulk of the parallactic motion.  The quantity $ X_{i0} $ is the
reference position of star $ i $ at time $ t_0 $, and $ X_{\rm ref,j}
$ is the offset of image $ j $ in the $ X $ direction---in essence,
the position of the reference scan line for image $ j $ on the sky.
The astrometric motion of star $ i $ in the $X$ direction is described
by the $ X $ component of the proper motion, $ pmx_i $, and the
parallax $ \pi_i $, applied with the epoch-dependent parallax factor $
f_j $.  The term $f_j$ is the projection (for unit parallax) of the
parallactic motion in the $X$ direction at the time of the
observations.  Note that the proper motion can only be relative, since
any change of all proper motions by the same amount can be subsumed
into a change of $ X_{\rm ref,j} $ for each epoch.  As far as the
astrometric model is concerned, parallaxes are also relative; however,
the degeneracy in the conversion to absolute parallaxes can be broken
by using spectrophotometric distance estimates for the stars in the
field.  Finally, the model position must be corrected for the relative
rotation and geometric distortion of image $ j $ with respect to the
reference image.  The rotation term on the sky is $ R_j \, Y_{i0} $,
where $ R_j $ is the rotation of image $ j $ and $ Y_{i0} $ is the
static relative position of star $ i $ in rectified coordinates along
the $ Y $ direction with respect to the center of the field.  (Since
typical rotations are of order $ 10^{-5} $, a measurement of $ Y_{i0}
$ with a precision of $ \sim 1 $~pixel will suffice.)  The polynomial
term is determined as part of the model-fitting procedure, typically
as a second- or third-degree polynomial $ P (X_{\rm det} , Y_{\rm det}
) $, where $ X_{\rm det} $ and $ Y_{\rm det} $ are detector
coordinates; the total correction is determined by evaluating the
polynomial at every location along the trail of star $ i $ in image $
j $ and averaging the result.  By convention, the rotation and
polynomial term apply to each frame in relation to the first in a set.

Again, the M35 calibration data provide good guidance, as they were
obtained at two orientations which differed by $ 180 \degr $.  These data
are useful for defining a family of polynomials with the smallest
number of terms needed to adequately account for variable distortion
(at the same orientation) or static and variable distortion (at flipped
orientation).

In the final step we need to convert the relative parallaxes into
absolute, thus yielding a geometric distance estimate.  This is
accomplished by estimating the reference star parallaxes from the
spectrophotometric absolute magnitude estimates that come from
multiband photometry and medium-resolution spectroscopy.  The distance
estimate of the target star will be insensitive to uncertainties in
the distance of the reference stars so long as the set contains
objects which are bright and distant (e.g., red giants).

For the SY Aur field (and for other Cepheid fields in progress), we
obtained direct imaging with {\it HST} during the scanning
observations and measured photometry of all reference stars in the UV
({\Filter{F275W}},{\Filter{F336W}}), Str\"{o}mgren
({\Filter{F410M}},{\Filter{F467M}},{\Filter{F547M}}), and broad-band
({\Filter{F850LP}}) systems.  To this photometry we added $ J, H, $
and $ K $-band photometry from the 2MASS survey to provide a set of up
to 9 bands of photometry from $ 0.2 $ to $ 2.2 \micron $.  All the
photometry was of high S/N, with the exception of {\Filter{F275W}}
where only a third of the stars yielded a measurement ($
{\Filter{F275W}} < 22.8 $ mag).  Missing or excluded photometry was
recorded for stars which suffered cosmic ray hits, suffered blending
in the 2MASS data (as identified with {\it HST} {\Filter{F850LP}}
imaging), and for half the field not covered by {\Filter{F410M}}
imaging.  In practice, the average number of bands with useful
measurements per star was between 6 and 7.

To estimate the spectroscopic parallax for each reference star, we
generated a sample of 29,000 mock stars in the direction of SY Aur
using the Besan{\c{c}}on galaxy model \citep [and references
therein]{robin86, robin03}.  The thick disk of the model has been
updated to better fit SDSS and 2MASS data (Robin 2013, private communication).

This sample of mock stars has a distribution of the four parameters
($\log g $, initial mass, metallicity, and extinction), as expected
from the Besan{\c{c}}on model along the sight line to SY Aur.  The
extinction to the edge of the Galaxy is defined from
\cite{schlegel98}.  For each mock star we select the stellar model
from the Padova isochrone tables \citep{bressan12} whose parameters
best match the mock parameters.
\footnote{The zeropoints of the Padova isochrones are themselves based
on geometric distances by the use of {\it Hipparcos} parallaxes of nearby
stars and the use of eclipsing binaries to measure masses, and thus
are unlikely to have a considerable, systematic error in distance
scale}  Each stellar model includes a determination of the absolute
magnitude (including the mock extinction appropriate for each band) of
that model star in each of the measured bands, thus producing 29,000
mock stellar models.  The comparison of each mock model to the
measurements of a reference star produces a distance estimate (and an
estimate of the 4 nuisance parameters) and a likelihood that the model
is good based on the size of the $\chi^2$ statistic between model and
data.

We independently determined the temperature and luminosity class of
the majority of the reference stars via medium-resolution optical
spectra compared to template spectra. The spectra were obtained with
the Kast double spectrograph (Miller \& Stone 1993) on the 3~m Shane
reflector at Lick Observatory and DIS on the APO 3.5~m. Standard procedures were used for
the data reduction.

In order to combine the photometric and spectroscopic results, we
treated the spectroscopic term as an additional contribution to the
$\chi^2$ statistic on the basis of the agreement between the
temperature and the $ \log g $ term given for the best-matching
spectral template, including both spectral and luminosity class.  The
spectral templates used were from \citep{pickles98} and the ELODIE
database \citep{prugniel01}.  This is especially important for
discriminating dwarfs from giants, for which the spectroscopic
contribution is often more powerful at discrimination than photometry.
Each star's distance (and expected parallax) was determined from the
normalized, summed product of mock distances and likelihoods (i.e.,
the Bayesian mean), with uncertainties fixed at 0.3 mag as derived
from Monte Carlo simulations of the models.  The parallax uncertainty
(i.e., systematic uncertainty in frame parallax) of the set of 31
fitted reference stars is $ 12 \muas $, well below our target
uncertainty.  As expected, most of this precision comes from the most
distant stars which are primarily red giants.  A single red giant at a
distance of $ 5 \kpc $ would give an uncertainty of $ 30 \muas $.
Indeed, the five most distant stars alone give an uncertainty in the
reduction to absolute parallax of $ 18 \muas $, equivalent to three
distant red giants.  Assuming a larger per-star uncertainty of 0.5 mag
increases the uncertainty in the constant parallax term to $ 21 \muas
$, still well below our overall accuracy for this field.  These
estimates allow us to break the degeneracy between relative parallaxes
and obtain an actual parallax estimate for each star, including the
Cepheid.

\subsubsection {Setting Up the Model}

In practice, the modeling process attempts to reproduce the relative
epoch-to-epoch position measurements for each star on the basis of
three parameters for each star: the position at the first epoch, the
parallax, and the proper motion.  One position and one proper motion
are fully degenerate; for simplicity, we assume that the mean position
at the first epoch and the mean proper motion of all stars considered
both vanish, but no result (except for a constant offset in the 
proper-motion terms) depends on these assumptions.  Another way to say this
is that we have no ability to establish a precise reference system for
either position or proper motion from the data available to us.  Each
epoch after the first is allowed a rotation, a constant term, and a 
second-degree polynomial
adjustment to match the first epoch; since there are about 34 stars
useful for measurement at each epoch, these additional 7 parameters
per epoch over which we marginalize do not place an undue burden on
the solution.

In addition, we use the {\it a priori} distance estimates based on
spectrophotometric parallaxes as priors for the parallax of the stars
in the field.  A prior is not used for the Cepheid, so that its
distance estimate is determined directly from its observed parallax.

The best values of the model parameters are determined using a version
of the Levenberg-Marquardt algorithm which includes proper error
propagation of the model uncertainties.  Among these parameters is the
{\it absolute} parallax of the target Cepheid, which then provides the
distance and error estimate for SY Aur.  A modest fraction of
the reference stars in the field are expected to be part of binaries 
with parameters that would cause a significant deviation from our
simple astrometric model.  This fraction depends on distance and
spectral class, but is $ \sim 10$--20\% for F and G stars 
at $ 1 \kpc $ on the basis of the distribution of binary properties in
\cite{duquennoy91}.  We run the global model iteratively after rejecting
outliers on the basis of their disproportionate contribution
to the total $ \chi^2 $; this results in the rejection of 8 reference stars.

\subsubsection {Multiparametric Model for SY Aurigae: Results}

The resulting fits for the SY Aur field are shown in Figure~7.  For
the remaining 26 reference stars, the spectroscopic parallax constraints are
shown, compared to the {\it a posteriori} parallax.  The proper-motion
estimates are subtracted from both data and model for ease of
examination.  For bright stars with long scans, the parallax
uncertainty is just under $ 40 \muas $, sufficient for a 16\%
measurement of parallax for two red giants, stars 4 and 10 at $ D=3.4 $
and $ 4.1 \kpc $, respectively.  The distance of the Cepheid SY Aur from
its parallax is $ D=2.3 \kpc $, making this star and others in the field the
most distant stars with well-determined parallax. This distance is in
good agreement with the expectation of $D \approx 2.1 \kpc $ based on the
Wesenheit {\PL} relation of Cepheids \citep{tammann03}.

The precision of the Cepheid parallax at this distance should be about
4\% based on optimal photon statistics, a designed
3600-pixel-long scan length, and position in the center of the field
away from larger and more uncertain distortions at the edge of the
field.  Because this particular Cepheid at $8.8 < V < 9.4$ mag is too
bright to avoid saturation in the deep scan when $V<10.6$ mag, its
measurement precision would be lower, about 6\%, owing to the reduced
photon statistics in the shallow scan.  However, the lack of bright
stars in this field (only 3 with astrometric errors of $ < 100 \muas $
in the shallow scan; see Fig.~6) degrades the precision of the transformation
between the shallow and deep scan for the Cepheid to between 50 and
$ 100 \muas $ and thus the precision of this parallax to 12\%.
Through better understanding of the parameters which affect the
parallax precision, we selected better Cepheid fields for new observations 
with more
reference stars, as well as filters and scan speeds which maximize
photon statistics, to get much closer to the attainable precision (see Fig.~6).  In
addition, the precision of the SY Aur parallax from these data should
improve through better empirical knowledge of the spatial
transformation between WFC3-UVIS filters garnered from ongoing spatial
scans of other Cepheid fields, so we consider this measurement to be
preliminary until collection of the full-sample of 19 Cepheids.

\section {Bright-Star Photometry}
     
\subsection{Spatial Scanning Photometry}
     
While ground-based observatories have imaged these bright MW
Cepheids in the NIR, systematic uncertainties in the flux
scale between the ground and {\it HST} photometric systems would limit the
precision of the Hubble constant independent of the measurement of
their parallaxes. The long-term internal stability of the nonstandard
{\it HST} photometric system in the NIR has been established to
$\sim$ 1\% \citep{kalirai11}. However, the NIR {\it HST} WFC3-IR
bandpasses of {\Filter{F160W}} and {\Filter{F125W}} used to observe Cepheids are not well
matched to ground-based bandpasses, which are set by natural breaks
in atmospheric OH emission and water transmission. The difference
between a typical ground-based $H$-band filter and its WFC3-IR
equivalent, {\Filter{F160W}}, is large, with color terms demonstrating a 20\%
difference between the measured brightness for stars with $J-H$
differing by 1 mag. In addition, ground-based systems suffer from
photometric instabilities at the few percent level owing to nightly and
hourly variations in the amount of precipitable water vapor and
aerosols in the atmosphere. The best-understood NIR ground-based
system, 2MASS, is calibrated to a precision of 0.02--0.03 mag
\citep{skrutskie06}.  Not surprisingly, systematic differences of 0.02
mag exist between the mean NIR magnitudes measured for {\it
the same Cepheids observed at different ground-based observatories},
even after accounting for the known differences in their photometric
systems \citep{monson11}.  These remaining differences likely reflect
the limitations with which the throughput of any ground-based system
is known in the NIR and present a critical {\it systematic}
uncertainty in relating ground-based NIR magnitudes of MW 
Cepheids to those in distant galaxies observed with {\it HST}.
     
While optical ground-based and {\it HST} systems are easier to 
cross-calibrate than those in the NIR, even the uncertainty between these
systems resulted in a reported 5\% systematic uncertainty in the
determination of H$_0$ by \cite{freedman01}.
     
The only way to ensure that future highest-quality Cepheid parallaxes from
{\it HST} or GAIA are fully leveraged is to observe the nearest MW 
Cepheids with {\it the same photometric system used to observe
their distant counterparts.}
However, it is challenging to accurately measure the brightness of
nearby, long-period Cepheids with {\it HST}. The 33 known MW
Cepheids with $P>10$~days and $D<5 \kpc $ have $3.5<H<7.5$~mag and
$6<V<10$ mag, and any would saturate in the shortest exposures possible
with WFC3-IR in the {\Filter{F125W}} and {\Filter{F160W}} filters.  In 
addition, the shortest possible exposure, 0.1 s with the smallest subarrays
(unsaturated for $H> 8 $ mag), limits the aperture radius to 32
pixels ($\sim 4\arcsec $), which is not ideal for distinguishing the
sky level from the wings of the PSF.  While it is possible to
accurately measure saturated sources by fitting to the unsaturated
wings of the light profile, it is important to minimize the degree of
saturation, as exclusive reliance on pixels progressively farther out
in the wings increases systematic uncertainties which propagate from
uncertain ratios between the peak and wing fluxes, including their
color dependence in broad bands.

One way to decrease the exposure time and the degree of saturation is
through spatial scanning during integration.  Spatial scanning of the
telescope reduces the effective exposure time a pixel ``sees'' a
source in proportion to the inverse of the scan speed.  The highest
scan speed available with {\it HST} (requiring gyroscope guiding) is 
$7\farcs8$ s$^{-1}$ which, for a pixel size of $ 0\farcs 13 $ for
WFC3-IR, is 0.017 s, or 0.03 s including flux from the off-peak
integration.  At this speed saturation begins at $H=7$ mag.  This is just under 
half the minimum integration of the smallest subarray allowed.  In the
optical the advantage of scanning bright targets is even greater, where
the $0\farcs 04$ pixel of WFC3-UVIS and the PSF produce a minimum
effective exposure time of 0.01 s, a factor of 50 shorter than the
minimum allowed exposure, a saturation limit of $V=7$ mag, and without
uncertainties associated with a variable time of shutter flight.  For
increasingly brighter targets above the saturation limit, successive
pairs of pixels adjacent to the peak saturate at a rate of
approximately one new pixel pair per magnitude, and information is lost
in blocks.  For optical photometry with WFC3-UVIS, \cite{gilliland10}
have shown that with a gain setting of 2 or higher, full-well
saturation occurs before digital saturation, so that saturated star
photometry is well measured by including the sum of the blooming
charge in the total without loss in precision or accuracy.

Another advantage of scanning bright stars instead of taking pointed 
images with subarrays comes from improved sampling of the
detector.  Pixel-to- pixel variations in the flat fields or positional
variation in quantum efficiency (QE) produces errors of about 0.01 mag
with the WFC3-IR detector \citep{riess11b}.  In addition, errors which
depend on pixel phase (e.g., resulting from imperfect knowledge of the
PSF) are reduced by scanning at an angle which varies the pixel phase,
yielding a result that is independent of pixel phase.

\subsection{SY Aurigae}

On JD 2,455,989 we obtained scanned observations of the MW
Cepheid SY Aur with WFC3-IR {\Filter{F160W}} at a commanded scan rate of
$ 7\farcs 5 / {\rm second} $.

We fit an empirical light profile (i.e., cross-section of a trailed
PSF) to each minirow (i.e., a 21 pixel row centered on the source).
The effective exposure time for each nondestructive sample of the
HgCdTe detector is a fixed number (determined from ground-based testing,
0.853 s to read out the $512 \times 512$ pixel subarray utilized) plus an
additional increment or decrement of exposure time to account for the
altered position of the moving source relative to the detector.  This
additional time is the standard interval between reads multiplied by
the fractional increase or decrease of the array that must be read out
before encountering the approaching or receding scan line (the sign of
which depends on the scan direction and whether the scan is above or
below the midline of the instrument which determines the readout
direction).\footnote{A simplified analogy relates the exposure time to
the time of flight for a ball traveling to a receiver who is moving
toward or away from the ball.}

The difference in integration between an approaching and receding scan
is quite large, amounting to an integration interval of 0.7092 s
over a length of 44 pixels between samples when scanning toward the
readout amplifier or 1.065 s over 65 pixels when scanning away
from the readout.  For each sample time interval, the count rate is
measured from the product of the mean minirow count in the sample and
the length of the sample scan divided by the time interval of the
sample.  The length divided by the interval is an empirical measure of
the scan rate, which was found to be about 4\% higher than requested.
After rejecting the central pixel of each minirow which is slightly
saturated, the measured count rate is 15.75 million $e^-$ s$^{-1}$ 
({\Filter{F160W}} = 6.706 mag) scanning toward and 15.49 million
$e^-$ s$^{-1}$ ({\Filter{F160W}} = 6.725 mag) scanning away.
This 2\% difference arises from edge effects causing us to overcount 
or undercount a partially filled pixel.  We take the average of $ 6.716
\pm 0.005$ mag to cancel the edge effects.  We increase this by 0.01 mag to $ 6.706
\pm 0.005$ mag to account for a 1\% systematic underestimate 
of bright sources in short exposures where trapping is seen to reduce the measured charge by 1\%.  
The row-to-row scatter is
3.25\% and appears to result from nonsmooth scanning.  It is worth
noting that a fractional error in the WFC3-IR sample time, if uniform
over all sample patterns (i.e., from the WFC3 clock running slow or
fast), would be negated in the measurement of relative distances
between MW and extragalactic Cepheids (when observed using the
same clock).  

Most known MW Cepheids have had their NIR light
curves measured by \cite{laney92} or \cite{monson11}.  Their periods
and phases have typically been determined to the third or fourth decimal
place from optical light curves, which assures that photometry from a
single epoch can be easily transformed to the mean magnitude with less
than 1\% statistical uncertainty (mean magnitudes are the standard
measure used for Cepheid distance determinations as shown in Figure~8).
For SY Aur, the expected difference between the single epoch and
the mean in {\Filter{F160W}} is 0.05 mag, for a final result of
${\Filter{F160W}}=6.67 \pm 0.01$ mag.  With $ E_{B-V} = 0.45$ mag and a 
period of 10.1 days, a Milky Way {\PL} relation \citep{benedict07,fouque07}
would give a distance modulus of $\mu_0 = 11.96 \pm 0.12$ mag, in good
agreement with the preliminary parallax distance modulus of $\mu_0 = 11.84$ 
mag measured here.

\section{Discussion}

We have presented a new approach to measuring high precision, relative
astrometry for stars with {\it HST} that can reach a final precision
of $ 20\hbox{--}40 \muas $ under optimal conditions.  We obtained
a preliminary measurement for the Cepheid SY Aurigae and several
reference stars in its immediate neighborhood; this is the first of a
planned 19 Galactic Cepheids to be measured with {\it HST} within the
next two years.

Measuring accurate parallaxes while simultaneously characterizing this
new observing mode has presented several new challenges.  Examples
include the impact of small variations in the telescope roll angle
during observations, the small but significant differences in the
geometric distortion solution as a function of filter and time, and
the difficulties involved in matching observations across filters in
order to tie the absolute parallax of the Cepheid target to that of a
sufficient number of reference stars.  

Some of these difficulties were
unknown or not fully appreciated at the start of this two-year
test program, the first of its kind, and have resulted in improvements
in the selection of target fields and observing modes for subsequent
targets.  For example, doubling the number of shallow scans improves
substantially the precision of the measurement, as shown in Figure~6.
Other changes could not be applied effectively in the middle of the
pilot campaign; for example, all targets selected for future
observations have been chosen to have more bright reference stars,
which yields an improved tie-in between target and reference stars,
and the depth of the shallow scan has been more carefully optimized.
These changes have improved the constraint on the Cepheid in the deep
frame to $ 30\hbox{--}40 \muas$, as shown for two new fields in
Figure~6.

For the remaining Cepheids, we have also acquired rapid
boustrophedonic or ``serpentine'' scans with the broad filter to serve
as an alternative to the narrow filter used for the shallow frame.
This observing mode can in principle yield an additional improvement
in the tie-in by a factor of the square root of the number of legs,
typically 4 or 5, and from the homogeneity of filters used in
transformations.

Lastly, we have found that we can improve on the polynomial
characterization of the variable distortion; in practice, the 5- or
even 9-parameter polynomial solutions appear to be restricted to a
smaller set of effective parameters, which in principle can be
identified and characterized through a principal-component analysis.
We expect this type of analysis can be substantially improved with the
availability of the statistics from the full program, including all
Cepheids and calibration observations, in order to better understand
and quantify the freedom required by the solution, and to improve our
knowledge of the mapping between {\it HST's} thermal state and
distortion.

The effort to measure parallaxes from space-based platforms for MW 
Cepheids has promise for anchoring a $ \sim 1\% $ determination of
the Hubble constant, an invaluable aid to cosmological investigations.
Spatial scanning astrometry with {\it HST} may also be suitable for a
much broader array of applications than considered here, including
exoplanet detection from astrometric motion or better relative
astrometry improving constraints on microlensing events. Although 
the precision possible with this technique rivals that of GAIA and
VLBI, the measurements are ultimately complementary because they can
be collected for different types of objects (VLBI) or with a different
class of systematics uncertainties (GAIA), to help ensure that parallax
measurements of stars beyond a kiloparsec are robust.

\bigskip 

This project was enabled by significant assistance from a wide variety
of sources.  We wish to thank Tom Harrison for his help determining
the luminosity class of stars from their spectra.  We thank Jeff
Silverman, Kelsey Clubb, Brad Cenko, Brad Tucker, and Waqas Bhatti for
their help obtaining and calibrating the spectra of reference stars in
the field of SY Aur; the staff at Lick Observatory also assisted.
Merle Reinhardt, George Chapman, William Januszewski, and Ken Sembach
provided help with the {\it HST} observations.  We thank Ed Nelan and
Fritz Benedict for productive discussions about the behavior of the
FGS.  We also thank Leo Girardi, Alessandro Bressan, Paola Marigo, for
the use of and assistance with their Padova isochrone database.
Support for this work was provided by NASA through programs GO-12679
and GO-13101 from the Space Telescope Science Institute, which is
operated by AURA, Inc., under NASA contract NAS 5-26555.  A.V.F. is
grateful for financial support from NSF grant AST-1211916, the TABASGO
Foundation, and the Christopher R. Redlich Fund.

\vfill
\eject

\bibliographystyle{apj}
\bibliography{mybibfile}

\begin {deluxetable}{cccccccc}
\tabletypesize{\footnotesize}
\tablecaption {Spatial scanning observations used in this paper}
\tablehead{
\colhead {Date} & \colhead {Rootname} & \colhead {Program ID}  & \colhead {EXPSTART} &
\colhead { Filter} & \colhead {Exp.~time} & \colhead {Scan rate} & \colhead {Scan length} \\
\colhead{} & \colhead {} & \colhead {} & \colhead {(MJD)} & \colhead {} & \colhead { (seconds)} & \colhead {($\arcsec/\rm second$)} & \colhead {($\arcsec$)} }
\tablecolumns{8}
\startdata
\hline
\sidehead {\bf SY Aurigae}
\hline
 2011-09-26   &   ibtq21gfq    &  12679  &   55830.87723571  & \Filter{F606W}  & 455  & 0.316  & 143.78 \\
 2011-09-26   &   ibtq21ghq    &  12679  &   55830.88444645  & \Filter{F673N}  & 348  & 0.414  & 144.07 \\
 2011-09-26   &   ibtq21gfq    &  12679  &   55830.87723571  & \Filter{F606W}  & 455  & 0.316  & 143.78 \\
 2012-03-02   &   ibx202j3q    &  12794  &   55988.55573488  & \Filter{F606W}  & 405  & 0.355  & 143.77 \\
 2012-03-02   &   ibx202j7q    &  12794  &   55988.56899858  & \Filter{F606W}  & 405  & 0.355  & 143.77 \\
 2012-03-02   &   ibx202jbq    &  12794  &   55988.58226266  & \Filter{F606W}  & 405  & 0.355  & 143.77 \\
 2012-03-02   &   ibtq24jkq    &  12679  &   55988.62925340  & \Filter{F606W}  & 455  & 0.316  & 143.78 \\
 2012-03-02   &   ibtq24jmq    &  12679  &   55988.63651007  & \Filter{F673N}  & 348  & 0.414  & 144.07 \\
 2012-09-16   &   ibtq25qcq    &  12679  &   56186.80378314  & \Filter{F673N}  & 380  & 0.379  & 144.02 \\
 2012-09-16   &   ibtq25qeq    &  12679  &   56186.80970907  & \Filter{F673N}  & 380  & 0.379  & 144.02 \\
 2012-09-16   &   ibtq25qaq    &  12679  &   56186.79652610  & \Filter{F606W}  & 455  & 0.316  & 143.78 \\
 2012-09-16   &   ibtq25qgq    &  12679  &   56186.81616721  & \Filter{F606W}  & 455  & 0.316  & 143.78 \\
 2013-03-03   &   ibtq26odq    &  12679  &   56354.53130985  & \Filter{F673N}  & 380  & 0.379  & 144.02 \\
 2013-03-03   &   ibtq26ofq    &  12679  &   56354.53723577  & \Filter{F673N}  & 380  & 0.379  & 144.02 \\
 2013-03-03   &   ibtq26obq    &  12679  &   56354.52405281  & \Filter{F606W}  & 455  & 0.316  & 143.78 \\
 2013-03-03   &   ibtq26ohq    &  12679  &   56354.54369429  & \Filter{F606W}  & 455  & 0.316  & 143.78 \\
 2013-08-25   &   ic4010ncq    &  13101  &   56529.04306131  & \Filter{F606W}  & 348  & 0.410  & 142.68 \\
 2013-08-25   &   ic4010neq    &  13101  &   56529.04884834  & \Filter{F673N}  & 348  & 0.410  & 142.68 \\
%2013-08-25   &   ic4010neq    &  13101  &   56529.04884834  & \Filter{F673N}  & 348  & 0.410  & 142.68 \\
\hline
\sidehead {\bf M35}
\hline
  2012-12-16   &  ic4001vjq    &  13101  &   56277.55560208  & \Filter{F606W}  & 350  & 0.410  & 143.50 \\
  2012-12-16   &  ic4001vlq    &  13101  &   56277.56118097  & \Filter{F606W}  & 350  & 0.410  & 143.50 \\
  2012-12-16   &  ic4001vnq    &  13101  &   56277.56675949  & \Filter{F606W}  & 350  & 0.410  & 143.50 \\
  2012-12-16   &  ic4001vpq    &  13101  &   56277.57233837  & \Filter{F606W}  & 350  & 0.410  & 143.50 \\
  2012-12-16   &  ic4001vrq    &  13101  &   56277.57791689  & \Filter{F606W}  & 350  & 0.410  & 143.50 \\
  2012-12-16   &  ic40a1vwq    &  13101  &   56277.62121541  & \Filter{F606W}  & 350  & 0.410  & 143.50 \\
  2012-12-16   &  ic40a1vyq    &  13101  &   56277.62679430  & \Filter{F606W}  & 350  & 0.410  & 143.50 \\
  2012-12-16   &  ic40a1w0q    &  13101  &   56277.63237282  & \Filter{F606W}  & 350  & 0.410  & 143.50 \\
  2012-12-16   &  ic40a1w2q    &  13101  &   56277.63795171  & \Filter{F606W}  & 350  & 0.410  & 143.50 \\
  2012-12-16   &  ic40a1w4q    &  13101  &   56277.64353023  & \Filter{F606W}  & 350  & 0.410  & 143.50 \\
  2012-12-17   &  ic4003bjq    &  13101  &   56278.22042837  & \Filter{F621M}  & 350  & 0.410  & 143.50 \\
  2012-12-17   &  ic4003blq    &  13101  &   56278.22600726  & \Filter{F621M}  & 350  & 0.410  & 143.50 \\
  2012-12-17   &  ic4003bnq    &  13101  &   56278.23158578  & \Filter{F621M}  & 350  & 0.410  & 143.50 \\
  2012-12-17   &  ic4003brq    &  13101  &   56278.23716467  & \Filter{F621M}  & 350  & 0.410  & 143.50 \\
  2012-12-17   &  ic4003btq    &  13101  &   56278.24274319  & \Filter{F621M}  & 350  & 0.410  & 143.50 \\
  2012-12-17   &  ic40a3c5q    &  13101  &   56278.28586837  & \Filter{F673N}  & 350  & 0.410  & 143.50 \\
  2012-12-17   &  ic40a3c7q    &  13101  &   56278.29144689  & \Filter{F673N}  & 350  & 0.410  & 143.50 \\
  2012-12-17   &  ic40a3c9q    &  13101  &   56278.29702578  & \Filter{F673N}  & 350  & 0.410  & 143.50 \\
  2012-12-17   &  ic40a3cbq    &  13101  &   56278.30260430  & \Filter{F673N}  & 350  & 0.410  & 143.50 \\
  2012-12-17   &  ic40a3cdq    &  13101  &   56278.30818319  & \Filter{F673N}  & 350  & 0.410  & 143.50 \\
  2012-12-24   &  ic4002aiq    &  13101  &   56285.66540541  & \Filter{F606W}  & 350  & 0.410  & 143.50 \\
  2012-12-24   &  ic4002akq    &  13101  &   56285.67098393  & \Filter{F606W}  & 350  & 0.410  & 143.50 \\
  2012-12-24   &  ic4002amq    &  13101  &   56285.67656282  & \Filter{F606W}  & 350  & 0.410  & 143.50 \\
  2012-12-24   &  ic4002aoq    &  13101  &   56285.68214134  & \Filter{F606W}  & 350  & 0.410  & 143.50 \\
  2012-12-24   &  ic4002aqq    &  13101  &   56285.68772023  & \Filter{F606W}  & 350  & 0.410  & 143.50 \\
  2012-12-24   &  ic40a2avq    &  13101  &   56285.73057912  & \Filter{F606W}  & 350  & 0.410  & 143.50 \\
  2012-12-24   &  ic40a2axq    &  13101  &   56285.73615763  & \Filter{F606W}  & 350  & 0.410  & 143.50 \\
  2012-12-24   &  ic40a2azq    &  13101  &   56285.74173652  & \Filter{F606W}  & 350  & 0.410  & 143.50 \\
  2012-12-24   &  ic40a2b1q    &  13101  &   56285.74731504  & \Filter{F606W}  & 350  & 0.410  & 143.50 \\
  2012-12-24   &  ic40a2b3q    &  13101  &   56285.75289356  & \Filter{F606W}  & 350  & 0.410  & 143.50 \\
  2013-01-03   &  ic4004dpq    &  13101  &   56295.50244302  & \Filter{F621M}  & 350  & 0.410  & 143.50 \\
  2013-01-03   &  ic4004drq    &  13101  &   56295.50802154  & \Filter{F621M}  & 350  & 0.410  & 143.50 \\
  2013-01-03   &  ic4004dtq    &  13101  &   56295.51360043  & \Filter{F621M}  & 350  & 0.410  & 143.50 \\
  2013-01-03   &  ic4004dvq    &  13101  &   56295.51917895  & \Filter{F621M}  & 350  & 0.410  & 143.50 \\
  2013-01-03   &  ic4004dxq    &  13101  &   56295.52475783  & \Filter{F621M}  & 350  & 0.410  & 143.50 \\
  2013-01-03   &  ic40a4e2q    &  13101  &   56295.56700302  & \Filter{F673N}  & 350  & 0.410  & 143.50 \\
  2013-01-03   &  ic40a4e4q    &  13101  &   56295.57258191  & \Filter{F673N}  & 350  & 0.410  & 143.50 \\
  2013-01-03   &  ic40a4e6q    &  13101  &   56295.57816043  & \Filter{F673N}  & 350  & 0.410  & 143.50 \\
  2013-01-03   &  ic40a4e8q    &  13101  &   56295.58373932  & \Filter{F673N}  & 350  & 0.410  & 143.50 \\
  2013-01-03   &  ic40a4eaq    &  13101  &   56295.58931783  & \Filter{F673N}  & 350  & 0.410  & 143.50 \\
\hline
\enddata
\end{deluxetable}

\begin{figure}[ht]
\vspace*{140mm}
\figurenum{1}
\includegraphics{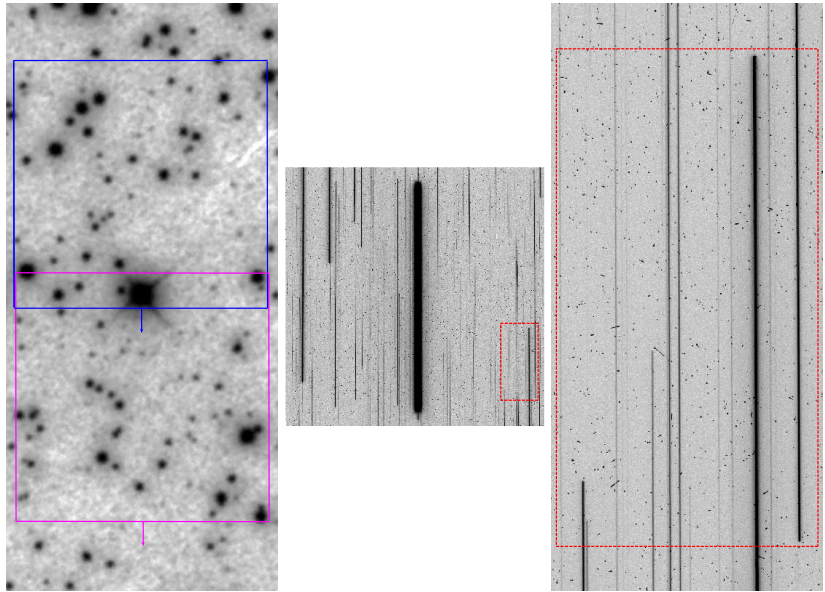}
\caption {(Left) Digital Sky Survey image of the field centered around
Cepheid SY Aur covered by WFC3-UVIS spatial scanning.  (Middle) Scan
image of the field in {\Filter{F606W}} from DD program 12879.  Points become
parallel lines with a greatly increased number of samples along the
scan direction and relative astrometry precision {\it perpendicular}
to the scan. (Right) Scan field marked with a box in the middle panel, 
magnified by a factor of 7.}
\end{figure}

\begin{figure}[ht]
\vspace*{140mm}
\figurenum{2}
\includegraphics{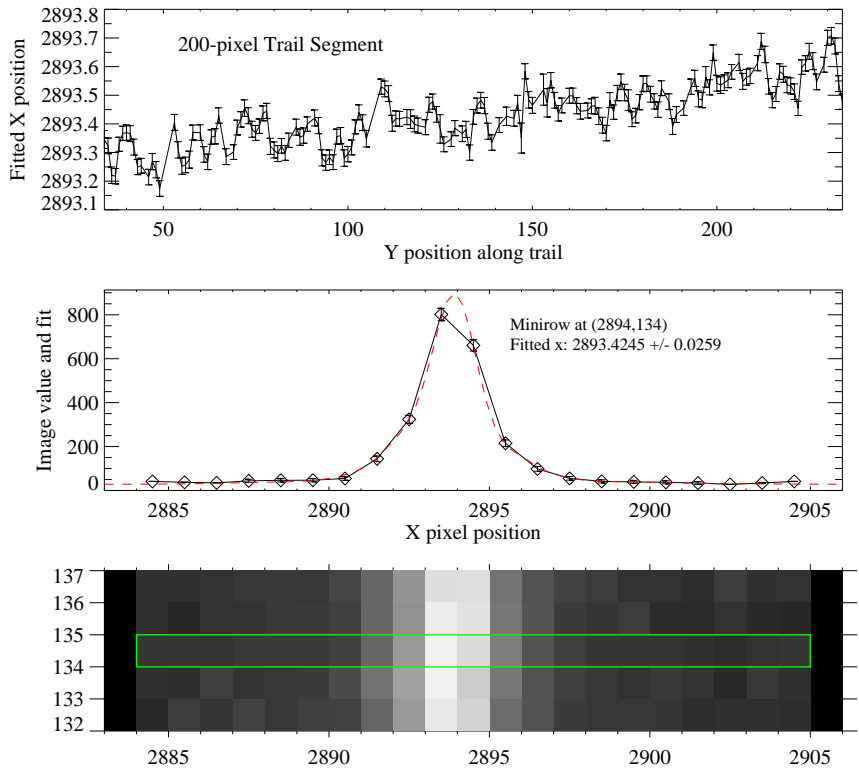}
\vspace {40pt}
\caption { The ``minirow'' unit of measure for spatial scans. 21-pixel 
rows centered on each scan line and their pixel values and
data quality flags (bottom) are used to measure the star position
along the row by fitting a section of a PSF (middle).  Measurement of
line positions in the detector $x$-coordinate reveal telescope jitter
(top).}
\end{figure}

\begin{figure}[ht]
\vspace*{140mm}
\figurenum{3}
\includegraphics{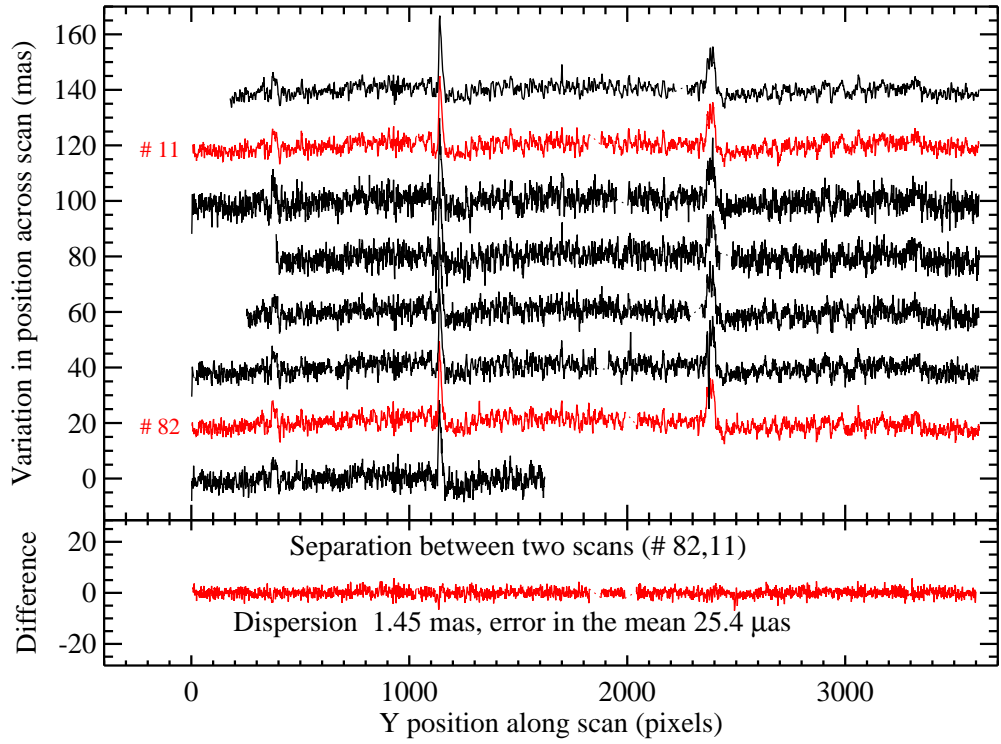}
\vspace {20pt}
\caption {(Top) Measured position of selected stars in the scanned
image of SY Aur, after subtraction of a low-order
polynomial. Correlated offsets indicate telescope jitter perpendicular
to the scan. (Bottom) Difference in position perpendicular to the
scan, the direction of parallax, for two stars.  The difference removes
the jitter noise which is correlated for each star and averages down
to a mean $ 25 \muas $ precision.}
\end{figure}

\begin{figure}[ht]
\vspace*{140mm}
\figurenum{4}
\includegraphics{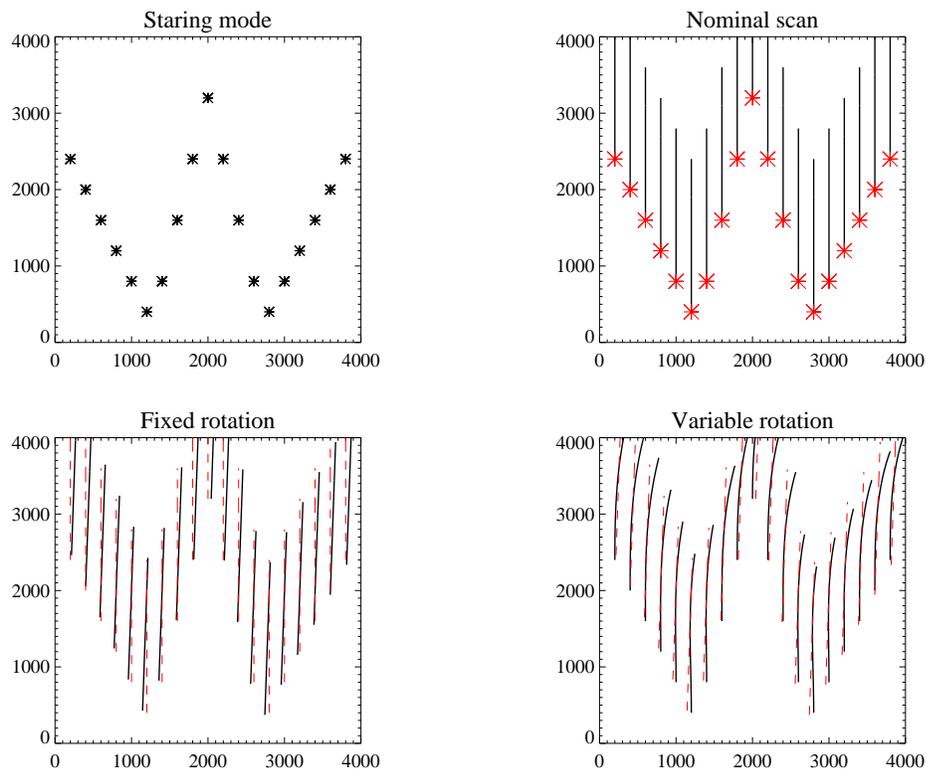}
\vspace {20pt}
\caption {Illustrations of the effect of static or variable rotation of
the telescope on lines during spatial scanning.  Fixed rotation and
variable rotation, the latter causing nonparallelism of scan lines,
must be measured and corrected from the rectified scan lines.}
\end{figure}

\begin{figure}[ht]
\vspace*{140mm}
\figurenum{5}
\includegraphics{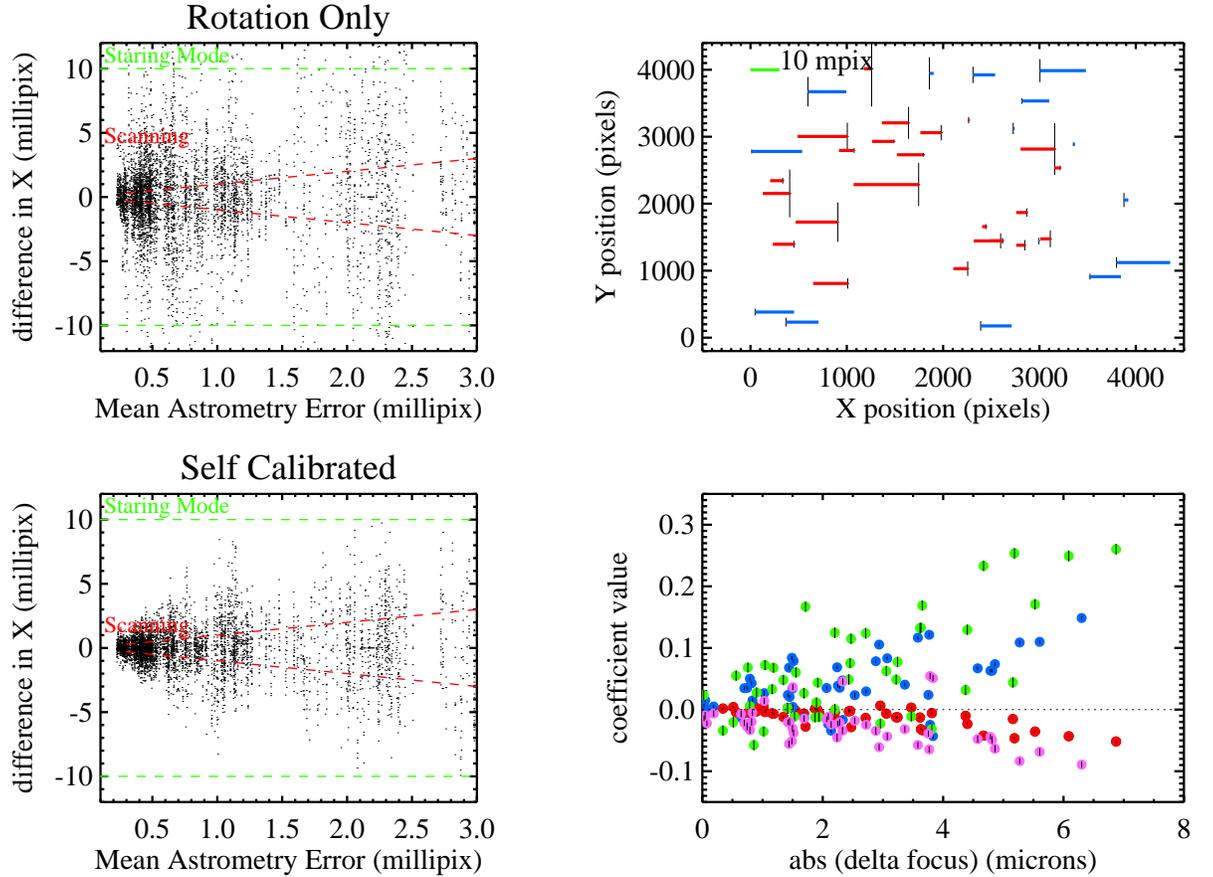}
\vspace {20pt}
\caption {Calibration of frame-to-frame variable distortion from
observations of stars in M35.  Excess dispersion in the relative
positions of stars over a continuous sequence of scans is apparent
(upper left).  Relating the position differences to the mean location
of each star on the detector reveals large-scale distortions.  A 
low-order polynomial can be used to measure and remove these
time-dependent distortions (lower left) to reach (or nearly reach) the 
photon statistics.  The polynomial coefficients appear to correlate with 
the modeled thermal state of {\it HST} and the modeled focus position
resulting from thermally induced piston motion of the secondary.}
\end{figure}

\begin{figure}[ht]
\vspace*{140mm}
\figurenum{6}
\includegraphics{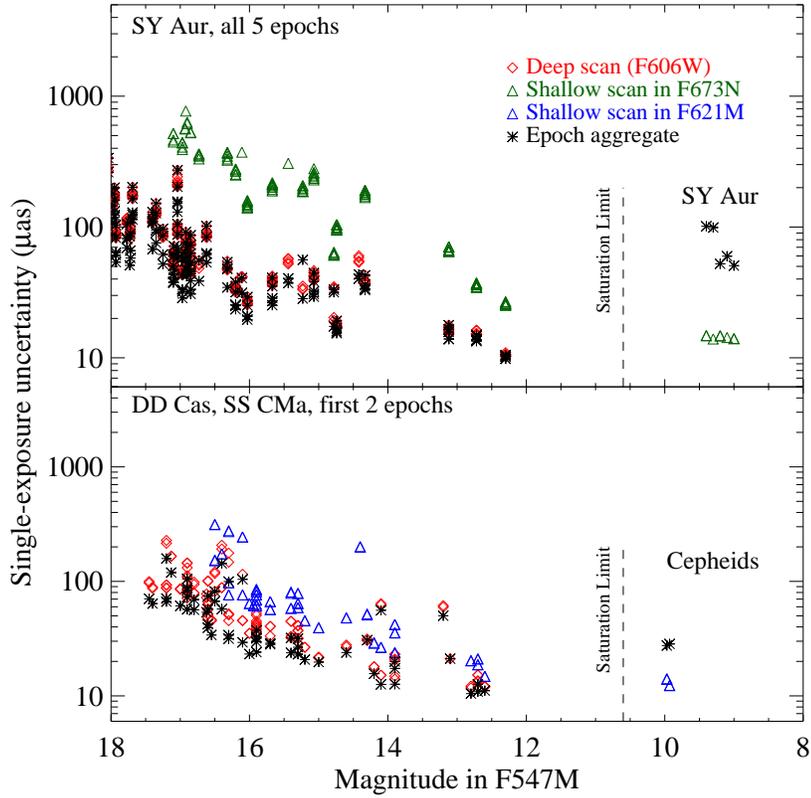}
\vspace {20pt}
\caption {Astrometric precision of star positions in spatial scans of
Cepheid fields.  A deep scan in {\Filter{F606W}} (shown in red) measures 
many stars to better than $ 40 \muas $ but saturates the Cepheid.  A
shallow scan in {\Filter{F673N}} (green) measures the Cepheid to $\sim 20 
\muas$, but the paucity of stars with measurements better than $ 100 \muas
$ reduces the precision of the modeled position of SY Aur in the deep
frame.  By choosing a field with more reference stars, using a more efficient
filter, and doubling the number of scans, we can measure Cepheids to
$ \sim 30 \muas $ (bottom panel).}
\end{figure}

\begin{figure}[ht]
\vspace*{180mm}
\figurenum{7}
\includegraphics{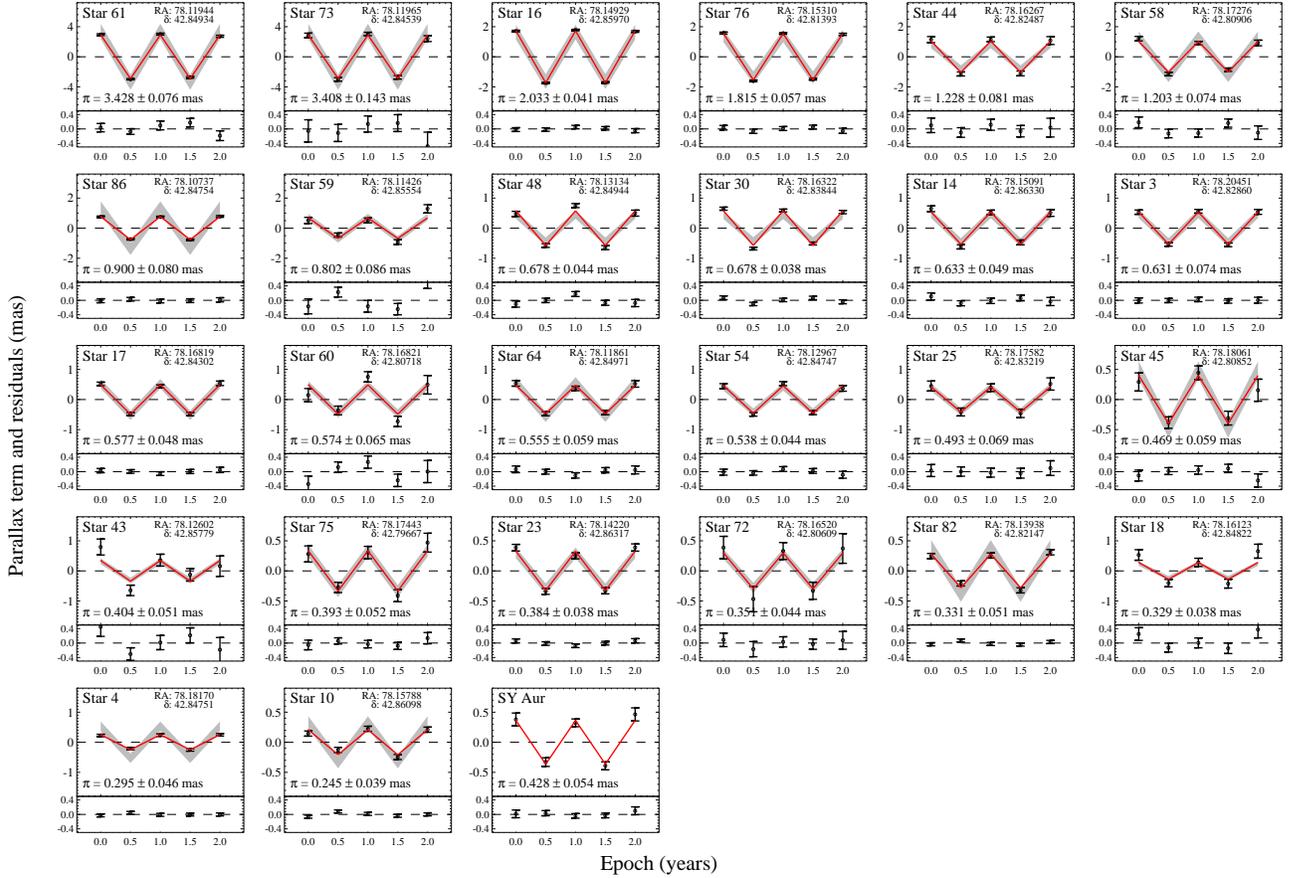}
\vspace {-40pt}
\caption {Individual stellar parallaxes in the field of SY Aur.
Ordered from closest to farthest are the measured parallax (red line)
and spectrophotometric parallax (grey band with $\pm 2\sigma$
width) used to reduce parallaxes from relative to absolute.  Fitted
proper motions have been subtracted from the measurements and fits for
ease of viewing.  Stars 4 and 10 are red giants with parallaxes
yielding $D=3.4 \pm 0.5$ and $4.1 \pm 0.6 \kpc $.  The Cepheid SY Aur
has a parallax yielding $D=2.3 \kpc$, in good agreement with an expectation
of $ 2.1 \kpc $ from the {\PL} relation.}
\end{figure}

\begin{figure}[ht]
\vspace*{140mm}
\figurenum{8}
\includegraphics{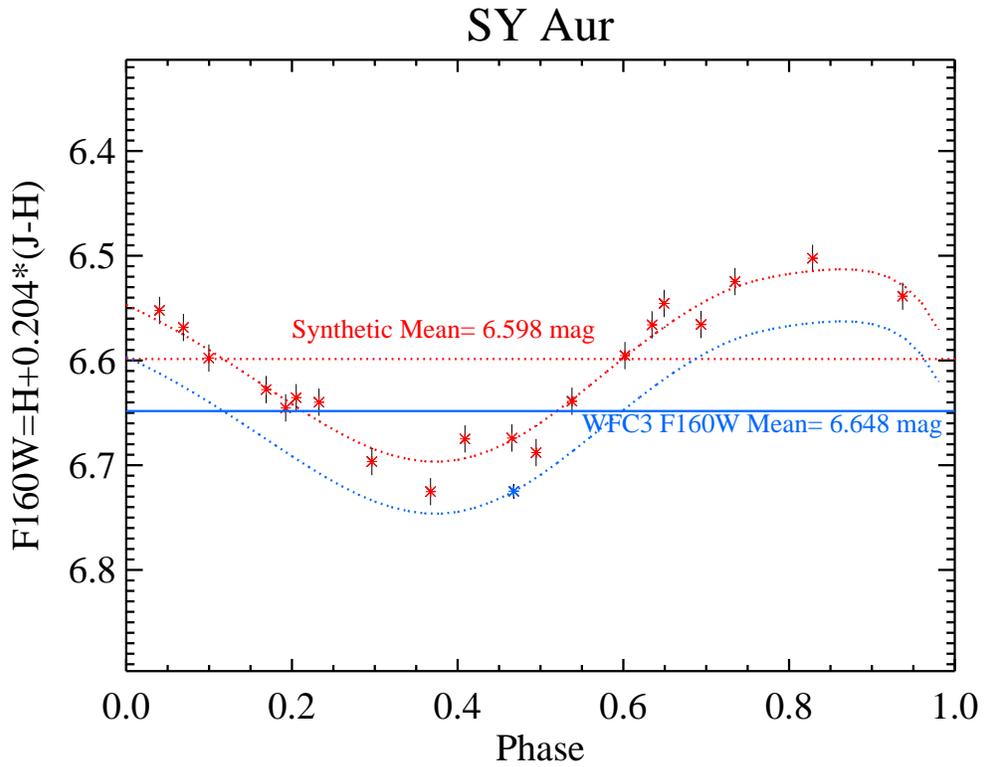}
\vspace {20pt}
\caption {Near-infrared spatial scan photometry of SY Aur.  A
photometric measurement with WFC3-UVIS {\Filter{F160W}} is transformed
from the indicated phase to the mean using the $J$ and $H$-band light
curves from \cite{monson11}.  The result is a measurement on the same
photometric system routinely used for extragalactic Cepheids at $ D >
20 \, \rm Mpc $.}
\end{figure}

\end{document}